\documentclass[onecolumn,draft]{IEEEtran}
\usepackage[mathscr]{eucal}
\usepackage[cmex10]{amsmath}
\usepackage{epsfig,epsf,psfrag}
\usepackage{amssymb,amsmath,amsthm,amsfonts,latexsym}
\usepackage{graphicx,bm,xcolor,url,overpic}
\usepackage{fixltx2e}
\usepackage{array}
\usepackage{verbatim}
\usepackage{bm}
\usepackage{algorithmic}
\usepackage{algorithm}
\usepackage{verbatim}
\usepackage{textcomp}
\usepackage{mathrsfs}
\usepackage{epstopdf}
\usepackage{mathtools}
\usepackage{pgfplots}
\usepackage{cite}
\usepackage{bbm}
\usepackage{color}
\usepackage{multirow,longtable}
\usepgfplotslibrary{fillbetween}

\newcommand{\nn}{\nonumber} 
\newcommand{\rk}{\mathrm{rk}}

\newtheorem{theorem}{Theorem} 
\newtheorem{lemma}[theorem]{Lemma}

\newtheorem{proposition}[theorem]{Proposition}

\newtheorem{definition}{Definition}

\newtheorem{remark}{Remark}

\usepackage{hyperref}
\hypersetup{colorlinks = true,
		linkcolor = blue,
		urlcolor = blue,
		citecolor = red,
		anchorcolor = green,}

\allowdisplaybreaks[1]

\begin{document}

\title{Characterizing Linear Memory-Rate Tradeoff of Coded Caching: The $(N,K)=(3,3)$ Case}
\author{\IEEEauthorblockN{Daming Cao and Yinfei Xu} 
\thanks{Daming Cao is with Department of Computer Science, National University of Singapore, Singapore (Email: dcscaod@nus.edu.sg). The work of D. Cao is partially completed in School of Information Science and Engineering at Southeast University during the PHD period.

Yinfei Xu is with School of Information Science and Engineering, Southeast University, China (Email: yinfeixu@seu.edu.cn).
}
}

\maketitle

\begin{abstract}
We consider the cache problem introduced by Maddah-ali and Niesen \cite{maddah2014fundamental} for the $(N,K)=(3,3)$ case, and use the computer-aided approach to derive the tight \textit{linear} memory-rate trade-off. Two lower bounds $10M+6R\geq 15$ and $5M+4R\geq 9$ are proved, which are non-Shannon type. A coded linear scheme of point $(M,R)=(0.6,1.5)$ is constructed with the help of symmetry reduction and brute-force search.

\end{abstract}

\IEEEpeerreviewmaketitle
\section{Introduction}

The coded cache problem in information theory is introduced by Maddah-Ali and Niesen \cite{maddah2014fundamental}. This framework has been extended to many scenarios and numerous results have been derived. Many of these extensions/results focus on the optimal memory-rate trade-off. Although various approaches have been proposed to study the  rate-memory trade-off, for the general case where the cached content can be coded, the optimal characterization of the trade-off remains open, except in some special cases, i.e., $N=K=2$ \cite{maddah2014fundamental}, $K=2$ and arbitrary $N$ \cite{tian2018symmetry,cao2019coded}, $K=3$ and $N=2$ \cite{tian2018symmetry}.

We revisit the framework in \cite{maddah2014fundamental} for the $(N,K)=(3,3)$ case. Instead of fully characterizing the optimal memory-rate trade-off, we derive a weaker  characterization where the cached content and the delivery messages are linear block codes. We use the computer-aided approach to prove both achievability and converse. For the converse, we combining two techniques, namely the computed-aided approach with symmetry reduction by Tian \cite{tian2018symmetry}, and the linear rank inequality with common information property by Hammer et al. \cite{hammer2000inequalities} and Dougherty et al. \cite{dougherty2009linear}, and derive two lower bounds which are non-Shannon type. On the other hand, for the achievability, we propose a symmetric structure for the cache placement, which significantly reduce the design complexity of both caches and delivery messages. Based on the numerical solutions from the computed-aided converse, an achievable coded cache placement with the symmetric structure is obtained by using the brute-force search.

\section{System Model and Existing Results}
\subsection{System Model}
We consider the cache problem introduced by Maddah-ali and Niesen \cite{maddah2014fundamental} for the $(N,K)=(3,3)$ case. For completeness, we briefly revisit the system model. The problem consider a system with one server connected to $K=3$ users through a shared, error-free link. The server has access to a database of $N=3$ independent equal-size files, each of size $F$ bits, denoted by $W_1$, $W_2$ and $W_3$. Each user is equipped with an equal-size local caches with capacities of $MF$ bits. The system operates in two phases. In the \textit{placement phase}, the users are given access to the entire database and fill their caches in an error-free manner. The contents of the caches after the placement phase are denoted by $Z_1$, $Z_2$ and $Z_3$, respectively. In the delivery phase, each user requests a single file from the server, where $d_k$ denotes the index of the file requested by User $k$, $k=1,2,3$. After receiving the demand pair $D \triangleq (d_1,d_2,d_3)$, the server generating a message of size $RF$ bits, denoted by $X^D$, and transmits the message over the shared channels to all users to satisfy their demands. 

Let $W_k$, $k=1,2,3$ be the independent random variables each uniformly distributed over $[2^F]$ for some $F\in \mathbbm{N}$. Then a \textit{$(M,R)$ cache scheme} for this system consists of:

\begin{enumerate}
    \item $K$ caching functions
\begin{align}
\phi_k: [2^{F}]^N \rightarrow [2^{MF}], 
\end{align}
which map the database into cache contents of the users, denoted by $Z_k=\phi_k(W_1, W_2, W_3)$, $k=1,2,3$.

\item $N^2$ encoding functions, one for each demand pair,
\begin{align}
f^D:  [2^{F}]^N \rightarrow [2^{RF}],  
\end{align}
  
\item $KN^2 $ decoding functions, one for each demand pair,
\begin{align}
g_k^D: [2^{MF}] \times [2^{RF}] \rightarrow [2^{F}], k=1,2,3, 
\end{align}
which decodes the desired file $W_{d_k}$ as $\hat{W}_{d_k}$ at User $k$ from the cached content at User $k$, the messages transmitted over the shared link.
\end{enumerate}
The probability of the cache scheme is defined as:
\begin{equation}
 P_{\mathrm{e}}=\max_{D\in[N]^K}\Pr\{(\hat{W}_{d_1},\hat{W}_{d_2},\hat{W}_{d_3})\neq (W_{d_1},W_{d_2},W_{d_3})\}.
\end{equation} 
For clarity, we adopt the zero-error decoding criterion. 

Given above definitions and setting, the \textit{memory-rate trade-off} is defined as follows.
\begin{definition}
A pair $(M,R)$ is {\em achievable} if for large enough file size $F$, there exists a $(M,R)$ cache scheme with zero error probability. The closure of the set of all (M,R)-achievable pair is called the memory-rate region and is denoted as $\mathcal{R}$. Then the {\em memory-rate trade-off} is defined as
\begin{equation}
R^*(M)\triangleq \inf\{R:(M,R)\in \mathcal{R}\}
\end{equation}
\end{definition}

Since most of the achievable $(M,R)$ cache schemes are composed of the \textit{linear block codes}, we are  interested in the \textit{linear} cache scheme.
\begin{definition}
A $(M,R)$ cache scheme is {\em linear} if all of the cached content and delivery messages are linear block codes.
\end{definition}

Similarly, we can define the \textit{linear} memory-rate trade-off as follow.
\begin{definition}
A pair $(M,R)$ is {\em linear}-achievable if for large enough file size $F$, there exists a $(M,R)$ {\em linear} cache scheme with zero error probability. The closure of the set of all (M,R)-linear-achievable pair is called the {\em linear} memory-rate region and is denoted as $\mathcal{R}_L$. Then the {\em linear} memory-rate trade-off is defined as
\begin{equation}
R^*_L(M)\triangleq \inf\{R:(M,R)\in \mathcal{R}_L\}
\end{equation}
\end{definition}

\subsection{Existing Results}
For the converse, by using the computational approach, Tian \cite{tian2018symmetry} provides the \textit{best} lower bound for $R^*(M)$ under the Shannon-type inequalities, i.e.,
\begin{equation}
\begin{cases}
3M+R^*(M)\geq 3\\
6M+3R^*(M)\geq 8\\
M+R^*(M)\geq 2\\
2M+3R^*(M)\geq 5\\
M+3R^*(M)\geq3
\end{cases} 
\end{equation}

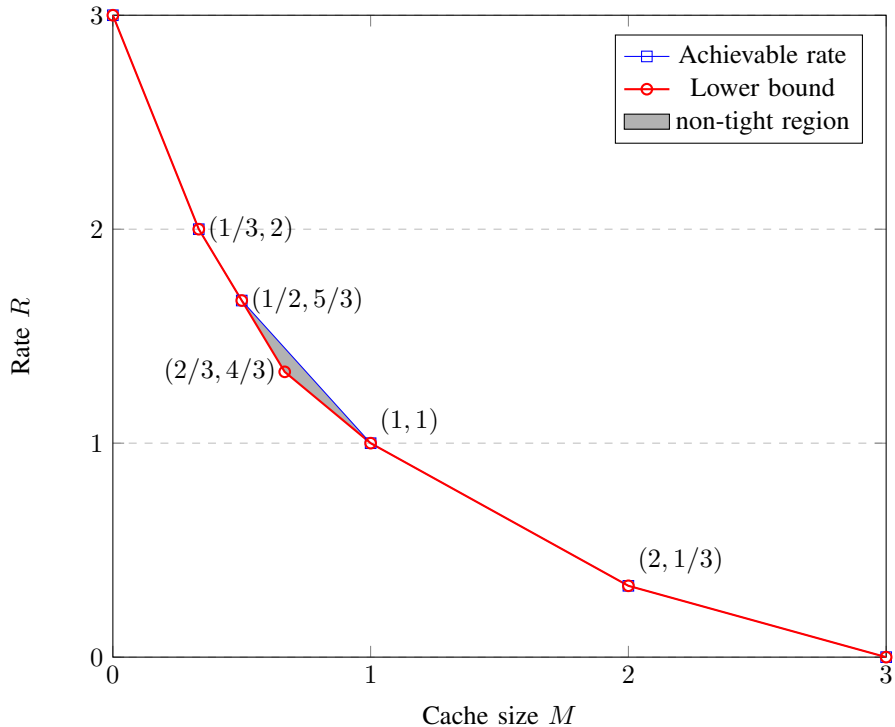
\begin{figure}[ht]
	\center
	\begin{tikzpicture}
	\center
	\begin{axis}[
	scale=1.5,
    xlabel={Cache size $M$},
    ylabel={Rate $R$},
    xmin=0, xmax=3,
    ymin=0, ymax=3,
    xtick={0,1,2,3},
    ytick={0,1,2,3},
    legend pos=north east,
    ymajorgrids=true,
    grid style=dashed,
    ]
    \addplot[
    name path=ach,
    color=blue,
    mark=square,
    ]
    coordinates {
    (0,3) (1/3,2)(1/2,5/3)(1,1)(2,1/3)(3,0)
    };
    \addplot[
    name path=conv,
    thick,
    color=red,
    mark=o,
    ]
    coordinates {
    (0,3) (1/3,2)(1/2,5/3)(2/3,4/3)(1,1)(2,1/3)(3,0)
    };
	\addplot [
        fill=gray!60, 
    ]
    fill between[
        of=ach and conv,
        soft clip={domain=1/2:1},
    ];   
    \legend{Achievable rate,Lower bound,non-tight region}
    \node[black,right] at (axis cs:1/3,2) {$(1/3,2)$};
    \node[black,right] at (axis cs:1/2,5/3) {$(1/2,5/3)$};
    \node[black,right] at (axis cs:1,1.1) {$(1,1)$};
    \node[black,right] at (axis cs:2,0.45) {$(2,1/3)$};
    \node[black,left] at (axis cs:2/3,4/3) {$(2/3,4/3)$};
    \end{axis}
	\end{tikzpicture}
	\caption{Existing rate-memory trade-off $R^*(M)$ for the $(3,3)$ cache problem.}
	\label{exist}
\end{figure}

For the achievability, the best known achievable pairs $(M,R)$ are proposed by several papers, i.e., $(1/3,2)$ in \cite{chen2016fundamental}, $(1/2,5/3)$ in \cite{gomez2016fundamental}, and $(1,1)$ and $(2,1/3)$ in \cite{maddah2014fundamental}. Note that all of these schemes are \textit{linear}. The existing results are shown in Fig.\ref{exist}

\section{Main Result}
\begin{figure}[ht]
	\center
	\begin{tikzpicture}
	\center
	\begin{axis}[
	scale=1.5,
    xlabel={Cache size $M$},
    ylabel={Rate $R$},
    xmin=1/3, xmax=1,
    ymin=1, ymax=2,
    xtick={0.333333333,0.5,0.66666666,1},
    xticklabels={1/3,1/2,2/3,1},
    ytick={1,1.33333333,1.6666666666,2},
    yticklabels={1,\Large $\frac{4}{3}$,\Large $\frac{5}{3}$,2},
    legend pos=north east,
    ymajorgrids=true,
    grid style=dashed,
    ]
    \addplot[
    name path=ach,
	very thick,
	dashdotted,]
    coordinates {
    (1/3,2)(1/2,5/3)(1,1)
    };

	\addplot[
    name path=newach,
    very thick,
    color=red,
    mark=o,]
    coordinates {
    (1/3,2)(1/2,5/3)(0.6,1.5)(1,1)
    };

    
    \addplot[
    name path=conv,
    thick,
    mark=square,]
    coordinates {
    (1/3,2)(1/2,5/3)(2/3,4/3)(1,1)
    };
    
   \legend{Known upper bound for $R^*(M)$,Tight characterization for $R^*_L(M)$, Known lower bound for $R^*(M)$}
   
    \node[black,right] at (axis cs:1/2,1.68) {$(1/2,5/3)$};
    \node[black,right] at (axis cs:0.63,1.6) {$(0.6,1.5)$};
    \node[black,left] at (axis cs:0.6,1.3) {$(2/3,4/3)$};
    \draw[->,>=stealth,very thick] (axis cs:0.63,1.6)--(axis cs:0.6,1.5);
    \draw[->,>=stealth,very thick] (axis cs:0.6,1.3)--(axis cs:2/3,4/3);
    \end{axis}
	\end{tikzpicture}
	\caption{Rate-memory trade-off $R^*(M)$ and $R^*_L(M)$ for the $(3,3)$ cache problem.}
	\label{linear}
\end{figure}
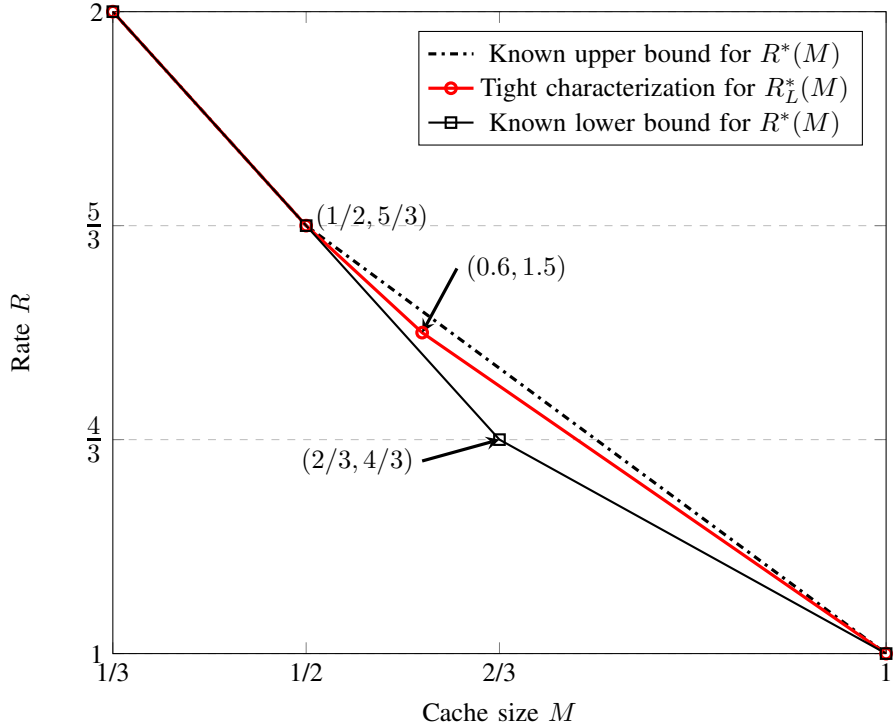

\begin{theorem}\label{TH1}
For the $(3,3)$ cache problem, the {\em linear} memory-rate trade-off $R^*_L(M)$ is fully characterized as follow 
\begin{equation}
R^*_L(M)=\max\left\{3-3M,\frac{8-6M}{3},\frac{15-10M}{6},\frac{9-5M}{4},\frac{5-2M}{3},\frac{3-M}{3}\right\}
\end{equation}
In other words, compared to the existing result (see Fig. \ref{linear}), the {\em linear} memory-rate trade-off $R^*_L(M)$ must additionally satisfy:
\begin{align}
10M+6R^*_L(M)&\geq 15\\
5M+4R^*_L(M)&\geq 9.
\end{align}
and the new corner point $(M,R)=(3/5,3/2)$ is {\em linear}-achievable. 
\end{theorem}


\section{Converse}
\subsection{Preliminaries}\label{Sec:conv-pre}
Before presenting the converse proof of Theorem \ref{TH1}, in this subsection, we briefly review the two techniques that we combine in this work, namely the computed-aided approach with symmetry reduction by Tian \cite{tian2018symmetry}, and the linear rank inequality with common information property by Hammer et al. \cite{hammer2000inequalities} and Dougherty et al. \cite{dougherty2009linear}.

The main idea in the computed-aided approach by Tian \cite{tian2018symmetry} is to use the information-theoretic inequality prover (ITIP) (or a linear programming (LP)) to identify the boundary of the memory-rate trade-off. However, the straightforward application can not work since the size of the linear programming is extremely large and is unbearable for the computer resource. Therefore, a critical step is to identify and formalize the symmetric structure and also to show the existence of optimal symmetric solutions. Subsequently, based on this symmetry property and problem setting, an equivalence relation for the entropy-quantity terms is given, which significantly reduces the size of the variables in LP and further make possible to use a symmetry-reduced LP with computer-aid in the cache problem. Furthermore, this equivalence relation can be described as follow: 
\begin{enumerate}
\item Symmetric rule: if two random variables terms satisfy some permutation constraint, these corresponding entropy quantities are equal, in other words, both quantities can be represent by a same variable in LP.
\item Decoding rule: if one random variable term can be decoded by the other random variable term, which means the corresponding conditional entropy is zero, both corresponding entropy quantities are equal.
\end{enumerate}

On the other hand, the key tool of the linear rank inequality with common information property by Hammer et al. \cite{hammer2000inequalities} and Dougherty et al. \cite{dougherty2009linear} is the common information property. More specific, a random variable $Z$ is a \textit{common information} of random variables $A$ and $B$ if it satisfies the following conditions:
\begin{align}
    H(Z|A)&=0,\\
    H(Z|B)&=0,\\
    H(Z)&=I(A;B).
\end{align}
Furthermore, if the random variables are generated/coming from a vector spaces, then the common information always exists. Subsequently, by introducing some new auxiliary random variables which are common information, the linear rank inequalities in \cite{hammer2000inequalities,dougherty2009linear} can be proved by Shannon-type inequalities even they are non-Shannon-type inequalities.
\subsection{Sketch of proof}
In the rest of this section, we present the converse proof of theorem \ref{TH1}. The proof follows by combining the techniques in \cite{tian2018symmetry} and \cite{hammer2000inequalities,dougherty2009linear} and is separated into three steps: 1) introduce two auxiliary random variables; 2) update the equivalence relation; 3) use the LP with symmetry reduction.

More specific, firstly, we introduce two auxiliary random variables $K_1$ and $K_2$, where $K_1$ is the common information of the random variables $Z_1X^{213}$ and $W_1$, and where $K_2$ is the common information of the random variables $W_1X^{123}$ and $W_2$. Since we consider the linear scheme, the variables $K_1$ and $K_2$ always exist. Secondly, based on the existing equivalence relation for the entropy-quantity terms without containing the auxiliary random variables, we only use the \textit{decoding rule} to update the equivalence relation. Clearly, this updated equivalence relation does not require any additional requirement for the optimal symmetric solutions. In other words, our steps do not break the optimality of the symmetric solutions. Finally, we use the  symmetry-reduced LP with computer-aid to obtain the low bound.

A detail of the equivalence relation is provided in Appendix \ref{apx:sym} and a ``checkable'' proof can be found in Appendix \ref{apx:conv}. 
\begin{remark}
It is noteworthy that this ``checkable'' proof may not be classical or standard since the equivalence relation is used in this proof. However, this non-classical part is equivalent or transformable to some standard converse techniques which uses the permutations to average the performance of all possible cases (cf. \cite{yu2017characterizing}). Furthermore, in some sense, this ``non-classical'' part is to use the permutations before giving solutions, while the standard ones are to use the permutations after giving solutions.
\end{remark}

\section{Achievability}
\subsection{Preliminaries}
For ease of notation, the three files are also denoted as $A$, $B$ and $C$, each of which is partitioned into ten subfiles of equal size,
denoted as $A_i$, $B_i$ and $C_i$, $i=1,2,\dots,10$, respectively. Refer to the linear block code, we construct the linear scheme in $\mathbf{GF}(2)$ and represent the schemes in a manner of information vectors and generated matrices. Specially, the information vectors is denoted by $\mathbf{W}=[A_1,A_2,\dots,A_{10},B_1,B_2,\dots,B_{10},C_1,C_2,\dots,C_{10}]$, and the generated matrices of the three files, the cached contents and the delivery messages are represented as the bold of the corresponding random variables. For example, the cached content of User 1 is the codeword vector $\mathbf{Z_1}\cdot \mathbf{W}^\mathrm{T}$, and the delivery message $X^{123}$ is the codeword vector $\mathbf{X^{123}}\cdot \mathbf{W}^\mathrm{T}$. In this section, we use $W_i$, $i=1,2,3$ and $A,B,C$, interchangeably, and also use the random variables and the corresponding codeword vectors interchangeably.    

Furthermore, note that these subfiles are independent and identical uniform distribution, therefore, the entropy of the random variable are the same (up to a constant factor $0.1F$) as the rank of the corresponding generated matrix. In the rest of this section, for ease of notation, we drop this constant normalized factor $0.1F$ for the entropies.

Subsequently, we introduce two column operations $f$ and $g$. For the operation $f$, it permutes the indexes of the columns as follow: 
\begin{equation}
    f:(1,2,\dots,30)\mapsto (21,22,\dots,30,1,2,\dots,20)
\end{equation}
or equivalently, it maps the notations of the files in the codeword vector as follow:
\begin{equation}
f:
\begin{cases}
A\mapsto B\\
B\mapsto C\\
C\mapsto A
\end{cases}.
\end{equation}
On the other hand, for the operation $g$, it permutes the indexes of the columns as follow: 
\begin{equation}
    g:(1,2,\dots,30)\mapsto (7,8,9,1,2,\dots,6,10,17,18,19,11,12,\dots,16,20,27,28,29,21,22,\dots,26,30)
\end{equation}
or equivalently, it maps the indexes of the subfiles for each files in the codeword vector as follow:
\begin{equation}
g:
\begin{cases}
i\mapsto i+3\quad(\mathrm{mod}\ 9) \qquad &1\leq i\leq 9\\
i\mapsto i &i=10 
\end{cases}
\end{equation}
For example, if the first component in the cache of User 1, i.e., $\mathbf{Z_1}[1]\cdot \mathbf{W}^\mathrm{T}$, is $A_2\oplus B_3\oplus C_9\oplus C_{10}$, then after processing on $\mathbf{Z_1}[1]$ by function $f\circ g$, we have $f\circ g(\mathbf{Z_1}[1])\cdot \mathbf{W}^\mathrm{T}=B_5\oplus C_6\oplus A_3\oplus A_{10}$. 

Moreover, we define the following mapping function $h$ :
\begin{equation}
h:
\begin{cases}
\mathbf{Z_1}\mapsto \mathbf{Z_2}\\
\mathbf{Z_2}\mapsto \mathbf{Z_3}\\
\mathbf{Z_3}\mapsto \mathbf{Z_1}\\
\mathbf{W_1}\mapsto \mathbf{W_2}\\
\mathbf{W_2}\mapsto \mathbf{W_3}\\
\mathbf{W_3}\mapsto \mathbf{W_1}
\end{cases}\quad \&\quad \begin{cases}
Z_1\mapsto Z_2\\
Z_2\mapsto Z_3\\
Z_3\mapsto Z_1\\
W_1\mapsto W_2\\
W_2\mapsto W_3\\
W_3\mapsto W_1\\
\end{cases}  
\end{equation}

Finally, we denote the set of the random variables which do not contain the delivery messages as the set $\mathcal{H}_Z$. Given a specific random variable in the set $\mathcal{H}_Z$, we define the vector which sequentially contains the element numbers of the files and the caches as the \textit{type} of this random variable. For example, the type of the random variable $Z_1W_1W_2$ is $(2,1)$, and the type of the random variable $W_2$ is $(1,0)$.

\subsection{Design the symmetric structure for the generated matrices of the caches}
Recall the new achievable pair is $(M,R)=(0.6,1.5)$, in other words, the entropy of each cache is at most $6$ and the entropy of each delivery message is at most $15$. 

Now let $\mathbf{Z_i}$ be a binary matrix with size $6\times 30$, $i=1,2,3$. Given an arbitrary assignment for the first two rows of 
$\mathbf{Z_1}$, in other words, $\mathbf{Z_1}[1-2]$ is arbitrarily given, we construct the rest rows in $\mathbf{Z_i}$, $i=1,2,3$ as follow:
\begin{align}
  \mathbf{Z_1}[3-4]&=f(\mathbf{Z_1}[1-2]),\label{eq:cons-z1_3}\\
  \mathbf{Z_1}[5-6]&=f(\mathbf{Z_1}[3-4]),\label{eq:cons-z1_5}\\
  \mathbf{Z_2}&=g(\mathbf{Z_1}),\label{eq:cons-z2}\\
  \mathbf{Z_3}&=g(\mathbf{Z_2}).\label{eq:cons-z3}
\end{align}

Note that the operators $f$ and $g$ are commutative and both composite operators $f^3$ and $g^3$ are identity operators, thus we have the following relations:
\begin{align}
    f(\mathbf{Z_i})&=\mathbf{Z_i},\quad i=1,2,3,\label{eq:fz}\\
    g(\mathbf{Z_i})&=h(\mathbf{Z_i}),\quad i=1,2,3\label{eq:gz}
\end{align}

Using the relations in \eqref{eq:fz} and \eqref{eq:gz}, we have the following proposition.
\begin{proposition}[Symmetry]\label{prop:sym}
For any two random variables in the set $\mathcal{H}_Z$, if they have the same type, then the corresponding entropies are equal.
\end{proposition}
\begin{IEEEproof}
Firstly, we consider the type $(1,1)$. Note that the operators $f$ and $g$ are the composition of some column interchange operations and recall the well-known fact that the column interchange operation does not change the rank of the matrix, thus we have the following conversions:
\begin{align}
    H(Z_i W_j)&=\rk\left(\begin{bmatrix}
    \mathbf{Z_i}\\
    \mathbf{W_j}
    \end{bmatrix}
    \right)=\rk\left(f\left(\begin{bmatrix}
    \mathbf{Z_i}\\
    \mathbf{W_j}
    \end{bmatrix}
    \right)\right)=\rk\left(\begin{bmatrix}
    f(\mathbf{Z_i})\\
    f(\mathbf{W_j})
    \end{bmatrix}
    \right)=\rk\left(\begin{bmatrix}
    \mathbf{Z_i}\\
    h(\mathbf{W_j})
    \end{bmatrix}
    \right)=H(Z_i h(W_j))\label{eq:heqf} \\
    H(Z_i W_j)&=\rk\left(\begin{bmatrix}
    \mathbf{Z_i}\\
    \mathbf{W_j}
    \end{bmatrix}
    \right)=\rk\left(g\left(\begin{bmatrix}
    \mathbf{Z_i}\\
    \mathbf{W_j}
    \end{bmatrix}
    \right)\right)=\rk\left(\begin{bmatrix}
    g(\mathbf{Z_i})\\
    g(\mathbf{W_j})
    \end{bmatrix}
    \right)=\rk\left(\begin{bmatrix}
    h(\mathbf{Z_i})\\
    \mathbf{W_j}
    \end{bmatrix}
    \right)=H(h(Z_i) W_j).  \label{eq:heqg} 
\end{align}
Therefore, by using the conversions \eqref{eq:heqf} and \eqref{eq:heqg}, the entropy of each random variable with type $(1,1)$ are identical. 

Similarly, we can derive the identical relationship for the rest types. 
\end{IEEEproof}

\begin{remark}
Proposition \ref{prop:sym} provides an equivalence relation for the entropy-quantity terms of the random variables in the set $\mathcal{H}_Z$. And this equivalence relation satisfies the symmetric rule introduced in the converse part (see Section \ref{Sec:conv-pre} and Appendix \ref{apx:sym}), in other words, the designed cache structure is symmetric (without considering the delivery messages). 
\end{remark}

\subsection{Brute-force search for the generated matrices of the caches}

Since the outer bound is obtained by ITIP, we have the numerical solutions for all entropy-quantity terms at the corner point $(0.5,1.6)$. However, even we only focus on the caches, in other words, we interest in the random variables in the set $\mathcal{H}_Z$, the optimal solutions are not unique. Therefore, to create coded multi-casting opportunities in the delivery phase as much as possible, intuitively, we choose an optimal solution which maximizes the sum of the entropies of the random variables in the set $\mathcal{H}_Z$ and provide it in Table \ref{tab:HV-cache}. Then the goal of this subsection is to find some cache constructions which match the table \ref{tab:HV-cache}.

\begin{table}[ht]
    \centering
    \caption{Entropy values at the corner point $(0.6,1.5)$}
    \begin{tabular}{c|c}
        \hline
        Entropy &  value \\
        \hline
           $H(Z_1)$  &  6\\
           $H(Z_1Z_2)$  &  12\\
           $H(Z_1Z_2Z_3)$  &  18\\
           $H(Z_1W_1)$  &  16\\
           $H(Z_1W_1W_2)$  &  24\\
           $H(Z_1Z_2W_1)$  &  22\\
           $H(Z_1Z_2W_1W_2)$  &  27\\
           $H(Z_1Z_2Z_3W_1)$  &  28\\
           $H(Z_1Z_2W_1W_2)$  &  30\\
           \hline
    \end{tabular}
    \label{tab:HV-cache}
\end{table}

Recall the construct of $Z_i$, $i=1,2,3$ in \eqref{eq:cons-z1_3} to  \eqref{eq:cons-z3} and note the observations $H(Z_1|W_1)=6$ and $H(Z_1|W_1W_2)=4$ from the table \ref{tab:HV-cache}, we may assume that the subfiles contained in the codeword vector $\mathbf{Z_1}[1-2]\cdot \mathbf{W}^\mathrm{T}$ are only from \textit{two} files. It is noteworthy that this encoding assumption dose not make any obvious contradiction to the entropy values in the table \ref{tab:HV-cache} (without entropy testing) and it reduces the design difficult, i.e., designing the linear combinations from two files rather three files. Without loss of generality, the codeword vector $\mathbf{Z_1}[1-2]\cdot \mathbf{W}^\mathrm{T}$ does not contain any subfiles in the file $C$.

Now we have reduced the size of all possible generated matrices $\mathbf{Z_i}$, $i=1,2,3$ from $2^{540}$ to $2^{40}$ based on the symmetric structure and the encoding assumption above. However, this size may still be too large and is unbearable/inefficient for the brute-force search (also for the manual design). Therefore, we further assume that the codeword vector $\mathbf{Z_1}[1-2]\cdot \mathbf{W}^\mathrm{T}$ only consider some partial subfiles, namely active encoding subfiles. Note the constraint $H(Z_1Z_2W_1W_2)=H(W_1W_2W_3)$, which means that each subfile at least appears once in some caches, therefore, we first consider a simple case that the active encoding subfiles of the codeword vector $\mathbf{Z_1}[1-2]\cdot \mathbf{W}^\mathrm{T}$ are the first three subfiles and the last subfiles, i.e., $(A_1,A_2,A_3,A_{10},B_1,B_2,B_3,B_{10})$. Then the size of all possible generated matrices $\mathbf{Z_i}$, $i=1,2,3$  is $2^{16}$ and it is acceptable for the brute-force search. Unfortunately, we do not find an achievable linear scheme\footnote{Due to the limit of coding ability, we can not conclude that there is no achievable linear scheme for this simple case.} even there exists some matched cache constructions. Thus, we slightly enlarge the active encoding subfiles of the codeword $\mathbf{Z_1}[1]\cdot \mathbf{W}^\mathrm{T}$ to the first four subfiles instead of the first three subfiles. Fortunately, we construct an achievable linear scheme and the corresponding cache construction is given in table \ref{tab:cache}.

\begin{table}[ht]
\caption{The cache construction for the corner point $(0.6,1.5)$.}
\begin{center}
\begin{tabular}{|c|c|c|c|}
    \hline
    \multirow{2}{*}{$Z_1$}& $A_1\oplus A_3\oplus A_4\oplus B_2\oplus B_{10}$ & $B_1\oplus B_3\oplus B_4\oplus C_2\oplus C_{10}$ & $C_1\oplus C_3\oplus C_4\oplus A_2\oplus A_{10}$\\
    \cline{2-4}
    & $A_2\oplus B_1$ & $B_2\oplus C_1$ & $C_2\oplus A_1$\\
    \hline
    \multirow{2}{*}{$Z_2$}& $A_4\oplus A_6\oplus A_7\oplus B_5\oplus B_{10}$ & $B_4\oplus B_6\oplus B_7\oplus C_5\oplus C_{10}$ & $C_4\oplus C_6\oplus C_7\oplus A_5\oplus A_{10}$\\
    \cline{2-4}
    & $A_5\oplus B_4$ & $B_5\oplus C_4$ & $C_5\oplus A_4$\\
    \hline
    \multirow{2}{*}{$Z_3$}& $A_7\oplus A_9\oplus A_1\oplus B_8\oplus B_{10}$ & $B_7\oplus B_9\oplus B_1\oplus C_8\oplus C_{10}$ & $C_7\oplus C_9\oplus C_1\oplus A_8\oplus A_{10}$\\
    \cline{2-4}
    & $A_8\oplus B_7$ & $B_8\oplus C_7$ & $C_8\oplus A_7$\\
    \hline
\end{tabular}\label{tab:cache}
\end{center}
\end{table}

\begin{remark}
In our construction, i.e., Table \ref{tab:cache}, the subfiles used in each cache are overlapped, e.g. $A_4,A_{10}$, which is a uncommon design for the case $K\cdot M \leq N$ and may also contradicts the intuitive design. 
\end{remark}

\subsection{Design the generated matrices of the delivery messages}
We partition all delivery messages into five parts in Table \ref{tab:sig-part} and prove Lemma \ref{lem:sig} that simplifies the design complexity.  

\begin{table}[ht]
\caption{The partition of all delivery messages.}
\begin{center}
\begin{tabular}{c|c}
    \hline
    Index & The demands of delivery messages \\
    \hline
    1 & $X^{AAA}$, $X^{BBB}$, $X^{CCC}$, \\
    \hline
    2 & $X^{ABC}$, $X^{BCA}$, $X^{CAB}$,\\
    \hline
    3 & $X^{ACB}$, $X^{CBA}$, $X^{BAC}$,\\
    \hline
    4 & $X^{ABB}$, $X^{BAB}$, $X^{BBA}$, $X^{BCC}$, $X^{CBC}$, $X^{CCB}$, $X^{CAA}$, $X^{ACA}$, $X^{AAC}$,\\
    \hline
    5 & $X^{ACC}$, $X^{CAC}$, $X^{CCA}$, $X^{CBB}$, $X^{BCB}$, $X^{BBC}$, $X^{BAA}$, $X^{ABA}$, $X^{AAB}$.\\
    \hline
\end{tabular}\label{tab:sig-part}
\end{center}
\end{table}

\begin{lemma}\label{lem:sig}
Given any generated matrices of the caches which satisfy the symmetric structure in \eqref{eq:cons-z1_3} to\eqref{eq:cons-z3}, in each part of Table \ref{tab:sig-part}, the generated matrices of the delivery messages are inter-transformable.
\end{lemma}
\begin{IEEEproof}
For Part 1 in Table \ref{tab:sig-part}, we give the transform mapping in \eqref{eq:AAA} and prove the achievability for the first mapping (right-arrow).
\begin{equation}\label{eq:AAA}
    \mathbf{X^{AAA}}\xrightarrow{\quad g\circ f \quad}
    \mathbf{X^{BBB}}\xrightarrow{\quad g\circ f \quad}
    \mathbf{X^{CCC}}\xrightarrow{\quad g\circ f \quad}
    \mathbf{X^{AAA}}
\end{equation}
Assume that the delivery message $X^{AAA}$ is achievable, in other words,
\begin{equation}
  H(W_1|X^{AAA}Z_i)=0,\quad i=1,2,3.  \label{eq:dec-w1}
\end{equation}
 Similarly to the proof of Proposition \ref{prop:sym}, we have the following conversion:
\begin{align}
    H(Z_i W_1 X^{AAA})&=\rk\left(\begin{bmatrix}
    \mathbf{Z_i}\\
    \mathbf{W_1}\\
    \mathbf{X^{AAA}}
    \end{bmatrix}
    \right)=\rk\left(g\circ f\left(\begin{bmatrix}
    \mathbf{Z_i}\\
    \mathbf{W_1}\\
    \mathbf{X^{AAA}}
    \end{bmatrix}
    \right)\right)\nn\\
    &=\rk\left(\begin{bmatrix}
    g\circ f(\mathbf{Z_i})\\
    g\circ f(\mathbf{W_1})\\
    g\circ f(\mathbf{X^{AAA}})
    \end{bmatrix}
    \right)=\rk\left(\begin{bmatrix}
    h(\mathbf{Z_i})\\
    \mathbf{W_2}\\
    \mathbf{X^{BBB}}
    \end{bmatrix}
    \right)=H(h(Z_i) W_2 X^{BBB})\label{eq:AAA-tr1} 
\end{align}
In a same way, we have
\begin{equation}
    H(Z_i X^{AAA})=H(h(Z_i) X^{BBB})  \label{eq:AAA-tr2}   
\end{equation}
Combining \eqref{eq:dec-w1}, \eqref{eq:AAA-tr1} and \eqref{eq:AAA-tr2}, we obtain that
\begin{equation}
  H(Z_i W_2 X^{BBB})=H(Z_i X^{BBB}),\quad i=1,2,3.
\end{equation}
Thus, we prove that the generated matrix $\mathbf{X^{BBB}}=g\circ f(\mathbf{X^{AAA}})$ is achievable. Similarly, the rest transform mapping is achievable. 

For the rest parts in Table \ref{tab:sig-part}, the transform mappings are provided in Appendix \ref{apx:sig-map} and the achievability can be proved in a similarly way. 
\end{IEEEproof}

Now, by using Lemma \ref{lem:sig}, we only need to design the generated matrices of the delivery messages $X^{AAA}$, $X^{ABC}$, $X^{ACB}$, $X^{ABB}$ and $X^{ACC}$. Obviously, the delivery message $X^{AAA}=A$ is always achievable since the rate is bigger than $1$.  

For the case $X^{ABC}$, motivated by the proof of Lemma \ref{lem:sig} and based on the observations that $g\circ f (\mathbf{Z_i})=h(\mathbf{Z_i})$ and $g\circ f (\mathbf{W_j})=h(\mathbf{W_j})$, we may hope that the matrix $\mathbf{X^{ABC}}$ satisfies the condition $\mathbf{X^{ABC}}=g\circ f (\mathbf{X^{ABC}})$. If so, we have 
\begin{equation}
    H(W_i|X^{ABC}Z_i)=H(h(W_i)|X^{ABC}h(Z_i)).
\end{equation}
Then we just need to guarantee that the User 1 can decode his required file $A$ from his cache $Z_1$ and the delivery message $X^{ABC}$. Furthermore, similarly to the symmetric structure of the cache generated matrices in \eqref{eq:cons-z1_3}-\eqref{eq:cons-z1_5}, we may further assume that the matrix $\mathbf{X^{ABC}}$ satisfies the following structure:
\begin{align}
    \mathbf{X^{ABC}}[6-10]&= g\circ f (\mathbf{X^{ABC}}[1-5])\\
    \mathbf{X^{ABC}}[11-15]&= g\circ f (\mathbf{X^{ABC}}[6-10]),
\end{align}
in other words, the construction work is reduced to design the first five rows $\mathbf{X^{ABC}}[1-5]$ instead of the whole matrix $\mathbf{X^{ABC}}$. Following the assumptions above, we find an achievable generated matrix $\mathbf{X^{ABC}}$ (similarly for $\mathbf{X^{ACB}}$) in table \ref{tab:sig-abc}. A checkable decoding processes are provided in Appendix \ref{apx:dec}.

\begin{table}[ht]
\caption{The delivery messages $X^{ABC}$ and $X^{ACB}$ for the corner point $(0.6,1.5)$.}
\begin{center}
\begin{tabular}{|c|c|c|}
    \hline
    \multicolumn{3}{|c|}{ABC}\\
    \hline
     $B_2\oplus B_{10}$ & $C_5\oplus C_{10}$ & $A_8\oplus A_{10}$\\
    \hline
     $A_4\oplus A_6\oplus A_7$ & $B_7\oplus B_9\oplus B_1$ & $C_1\oplus C_3\oplus C_4$\\
    \hline
     $A_7$ & $B_1$ & $C_4$\\
    \hline
     $A_5$ & $B_8$ & $C_2$\\
    \hline
     $A_1\oplus A_9\oplus B_2\oplus C_1$ & $B_3\oplus B_4\oplus C_5\oplus A_4$ & $C_6\oplus C_7\oplus A_8\oplus B_7$\\
    \hline
    \multicolumn{3}{|c|}{ACB}\\
    \hline
     $A_5\oplus A_{10}$ & $B_2\oplus B_{10}$ & $C_8\oplus C_{10}$\\
    \hline
     $A_7\oplus A_9\oplus A_1$ & $B_4\oplus B_6\oplus B_7$ & $C_1\oplus C_3\oplus C_4$\\
    \hline
     $A_4$ & $B_1$ & $C_7$\\
    \hline
     $A_8$ & $B_5$ & $C_2$\\
    \hline
     $A_6\oplus A_7\oplus B_2\oplus C_1$ & $B_3\oplus B_4\oplus C_8\oplus A_7$ & $C_9\oplus C_1\oplus A_5\oplus B_4$\\
    \hline
\end{tabular}\label{tab:sig-abc}
\end{center}
\end{table}

For the rest cases $X^{ABB}$ and $X^{ACC}$, although we do not have a similar symmetric structure for the generated matrices, we can still partially reduce the design complexity in a similar manner, which is benefited from the symmetric structure of the caches. The construction of the delivery messages $X^{ABB}$ and $X^{ACC}$ are given in Table \ref{tab:sig-abb} and a checkable decoding processes are provided in Appendix \ref{apx:dec}.

\begin{table}[ht]
\caption{The delivery messages $X^{ABB}$ and $X^{ACC}$ for the corner point $(0.6,1.5)$.}
\begin{center}
\begin{tabular}{|c|c|c|}
    \hline
    \multicolumn{3}{|c|}{ABB}\\
    \hline
     $B_2\oplus B_{10}$ & $A_4\oplus A_6\oplus A_7$ & $A_7\oplus A_9\oplus A_1$\\
    \hline
     $B_3\oplus B_4$ & $A_5\oplus C_{10}\oplus A_4$ & $A_8\oplus C_{10}\oplus A_7$\\
    \hline
     $A_{10}$ & $B_{10}$ & $C_{10}$\\
    \hline
     $B_1$ & $A_5$ & $A_8$\\
    \hline
    $B_4\oplus B_7$ & $B_6\oplus B_9$ & $B_5\oplus B_8$\\
    \hline
    \multicolumn{3}{|c|}{ACC}\\
    \hline
     $C_1\oplus C_3\oplus C_4$ & $A_5\oplus A_{10}$ & $A_8\oplus A_{10}$\\
    \hline
     $C_1\oplus C_2$ & $A_6\oplus A_7$ & $A_9\oplus A_1$\\
    \hline
     $A_{10}$ & $B_{10}$ & $C_{10}$\\
    \hline
     $C_2$ & $A_4$ & $A_7$\\
    \hline
     $C_5\oplus C_8$ & $C_4\oplus C_7$ & $C_6\oplus C_9$\\
    \hline
\end{tabular}\label{tab:sig-abb}
\end{center}
\end{table}

\begin{remark}
The delivery message $X^{ABB}$ gives a ``contradiction'' to an intuitive guess that if a file is not required in some demands, the corresponding delivery messages are independent of this file. 
\end{remark}

\newpage
\appendix
\subsection{The symmetry introduced in \cite{tian2016symmetry}}\label{apx:sym}
As we use it in the subsequent proof, for the completeness, we briefly restate the symmetry by Tian \cite{tian2016symmetry}. 

Let $\bar{\pi}(\cdot)$ and $\hat{\pi}(\cdot)$ be two permutation functions on the index set $\{1,2,3\}$, $\mathcal{Z}$ be a subset of $\{Z_1,Z_2,Z_3\}$, $\mathcal{W}$  be a subset of $\{W_1,W_2,W_3\}$, and $\mathcal{X}$ be a subset of {$\{X^D:D\in [3]\times[3]\}$}. Define the following operations:
\begin{align}
    \bar{\pi}\circ \hat{\pi}(\mathcal{W})&= \{W_{\hat{\pi}(i)}:W_i\in \mathcal{W}\} \label{eq:per_w}\\  
    \bar{\pi}\circ \hat{\pi}(\mathcal{Z})&= \{Z_{\bar{\pi}(i)}:Z_i\in \mathcal{Z}\} \\
    \bar{\pi}\circ \hat{\pi}(\mathcal{X})&= \left\{X^{\big(\bar{\pi}^{-1}(\hat{\pi}(d_1)),\bar{\pi}^{-1}(\hat{\pi}(d_2)),\bar{\pi}^{-1}(\hat{\pi}(d_3)\big)}:X^{(d_1,d_2,d_3)}\in \mathcal{X}\right\}\label{eq:per_x}
\end{align}
Now, the symmetry can be represented as
\begin{equation}
    H(\mathcal{W},\mathcal{Z},\mathcal{X})= H\big(\bar{\pi}\circ \hat{\pi}(\mathcal{W}),\bar{\pi}\circ \hat{\pi}(\mathcal{Z}),\bar{\pi}\circ \hat{\pi}(\mathcal{X})\big)\qquad \forall (\bar{\pi}(\cdot),\hat{\pi}(\cdot))\label{eq:h_sym}
\end{equation}

\subsection{The ``checkable'' converse proof of Theorem \ref{TH1}}\label{apx:conv}

Firstly, we specify the equivalence relation as follow:
\begin{enumerate}
    \item Symmetric rule: it is the equation \eqref{eq:h_sym} and is only available for the random variables without containing the auxiliary random variables $K_1$ and $K_2$. Moreover, the permutation is represented by one-line notation, and this equivalence relation is represented by the right arrow.
    \item Decoding rule: it follows the following forms/cases: 
    \begin{align}
        H(\cdot|W_1,W_2,W_3,\cdot)&=0\\
        H(W_{d_i}|X^D,Z_i,\cdot)&=0\\
        \left\{
        \begin{aligned}
            H(K_1|Z_1,X^{213},\cdot)&=0,\\
            H(K_1|W_1,\cdot)&=0,\\
            H(K_2|W_1,X^{123},\cdot)&=0,\\
            H(K_2|W_2,\cdot)&=0,    
        \end{aligned}
        \right.
    \end{align}
    Moreover, this equivalence relation is represented by the equal sign.
\end{enumerate}
For example, $H(W_1W_2Z_1)$ is equivalent to $H(W_1W_2Z_3K_1K_2)$ through the following relation:
\begin{equation}
    H(W_1W_2Z_1)\rightarrow H(W_1W_2Z_3) = H(W_1W_2Z_3K_1K_2)\qquad ((1,2,3),(3,2,1))  
\end{equation}
where the brackets is the corresponding permutation pair, i.e., $(\bar{\pi}(\cdot),\hat{\pi}(\cdot))=((1,2,3),(3,2,1))$. 

Now, we show the proof of the lower bound $10M+6R\geq 15$. For ease of checking, we present the equivalence relation for the entropy terms and the original Shannon-type inequality under each inequality. Based on the results of the computer-aided LP, we have:

\begin{align}
&H(W_1W_2W_3)-H(W_1W_2Z_1)-H(W_2W_3Z_3K_1K_2)+H(W_2Z_3K_1K_2)\leq 0\label{eq:conv1_beg}\\
&\qquad H(W_1W_2Z_1)\rightarrow H(W_1W_2Z_3) = H(W_1W_2Z_3K_1K_2)\qquad ((1,2,3),(3,2,1)) \nn\\
&\qquad \Leftrightarrow I(W_1:W_3|W_2Z_3K_1K_2)\geq 0\nn\\
&3H(W_1W_2W_3)-3H(W_1W_2Z_1Z_2)-3H(W_1W_2Z_1X^{123})+3H(W_1W_2Z_1)\leq 0\\
&\qquad H(W_1W_2W_3)=H(W_1W_2Z_1Z_3X^{123})\nn\\
&\qquad H(W_1W_2Z_1Z_2)\rightarrow H(W_1W_2Z_1Z_3)\qquad ((1,2,3),(1,3,2))\nn\\
&\qquad \Leftrightarrow I(Z_3;X^{123}|W_1W_2Z_1)\geq 0\nn\\
&H(W_1W_2Z_1Z_2X^{123})-H(W_1Z_1X^{123})-H(W_2W_3Z_3X^{213}K_1K_2)+H(W_2X^{213}K_1K_2)\leq 0\\
&\qquad H(W_1W_2Z_1Z_2X^{123})\rightarrow H(W_2W_3Z_1Z_3X^{213})= H(W_2W_3Z_1Z_3X^{213}K_1K_2)\qquad ((2,3,1),(1,3,2))\nn\\
&\qquad H(W_1Z_1X^{123})\rightarrow H(W_2Z_1X^{213})= H(W_2Z_1X^{213}K_1K_2)\qquad ((2,3,1),(1,3,2)) \nn\\
&\qquad \Leftrightarrow I(W_3Z_3;Z_1|W_2X^{213}K_1K_2)\geq 0\nn\\
&5H(W_1W_2Z_1Z_2X^{123})-5H(W_1W_2Z_1X^{123})-5H(W_1Z_1X^{123})+5H(W_1X^{123})\leq 0\\
&\qquad H(W_1W_2Z_1X^{123})\rightarrow H(W_1W_2Z_2X^{123})\qquad ((2,1,3),(2,1,3))\nn\\
&\qquad \Leftrightarrow I(Z_1;W_2Z_2|W_1X^{123})\geq 0\nn\\
&4H(W_1W_2Z_1X^{123})-4H(W_1Z_1)-4H(W_1X^{123})+4H(W_1)\leq 0\\
&\qquad H(W_1W_2Z_1X^{123})\rightarrow H(W_1W_2Z_2X^{123})= H(W_1Z_2X^{123})\qquad ((2,1,3),(2,1,3))\nn\\
&\qquad H(W_1Z_1)\rightarrow H(W_1Z_2) \qquad ((1,2,3),(2,1,3))\nn\\
&\qquad \Leftrightarrow I(Z_2;X^{123}|W_1)\geq 0\nn\\
&H(W_1W_2Z_1X^{123})-H(W_1Z_1X^{123})-H(W_1X^{123})+H(X^{213}K_1)\leq 0\\
&\qquad H(W_1W_2Z_1X^{123})\rightarrow H(W_1W_2Z_1X^{213})= H(W_1W_2Z_1X^{213}K_1)\qquad ((2,1,3),(1,2,3))\nn\\
&\qquad H(W_1Z_1X^{123})\rightarrow H(W_2Z_1X^{213})= H(W_2Z_1X^{213}K_1)\qquad ((2,3,1),(1,3,2))\nn\\
&\qquad H(W_1X^{123})=H(W_1X^{123}K_1)\nn\\
&\qquad H(W_1X^{123})\rightarrow H(W_1X^{213})= H(W_1X^{213}K_1)\qquad ((1,3,2),(2,3,1))\nn\\
&\qquad \Leftrightarrow I(W_1;W_2Z_1|X^{213}K_1)\geq 0\nn\\
&H(W_1W_2Z_1X^{123})-H(W_1W_2Z_1)-H(W_3Z_3X^{213}K_1)+H(W_3Z_3K_1)\leq 0\\
&\qquad H(W_1W_2Z_1X^{123})\rightarrow H(W_1W_3Z_3X^{213})= H(W_1W_3Z_3X^{213}K_1)\qquad ((3,1,2),(3,2,1))\nn\\
&\qquad H(W_1W_2Z_1)\rightarrow H(W_1W_3Z_3)= H(W_1W_3Z_3K_1)\qquad ((1,3,2),(3,1,2))\nn\\
&\qquad \Leftrightarrow I(W_1;X^{213}|W_3Z_3K_1)\geq 0\nn\\
&H(W_1W_2)-H(W_1)-H(W_2K_1K_2)+H(K_1)\leq 0\\
&\qquad H(W_1W_2)= H(W_1W_2K_1K_2)\nn\\
&\qquad H(W_1)= H(W_1K_1)\nn\\
&\qquad \Leftrightarrow I(W_1;W_2K_2|K_1)\geq 0\nn\\
&7H(W_1Z_1X^{123})-7H(Z_1)-7H(X^{123})\leq 0\\
&\qquad H(W_1Z_1X^{123})=H(Z_1X^{123})\nn\\
&\qquad \Leftrightarrow I(Z_1;X^{123})\geq 0\nn\\
&3H(W_1Z_1)-3H(W_1)-3H(Z_1)\leq 0\\
&\qquad \Leftrightarrow I(W_1;Z_1)\geq 0\nn\\
&H(W_2W_3Z_3X^{213}K_1K_2)-H(W_2Z_3K_1K_2)-H(W_2X^{213}K_1K_2)+H(W_2K_1K_2)\leq 0\\
&\qquad H(W_2W_3Z_3X^{213}K_1K_2)=H(W_2Z_3X^{213}K_1K_2)\nn\\
&\qquad \Leftrightarrow I(Z_3;X^{213}|W_2K_1K_2)\geq 0\nn\\
&H(W_3Z_3X^{213}K_1)-H(X^{213}K_1)-H(W_1Z_1X^{123})+H(X^{123})\leq 0\\
&\qquad H(W_1Z_1X^{123})\rightarrow H(W_3Z_3X^{213})\qquad ((3,2,1),(3,1,2))\nn\\
&\qquad H(X^{123})\rightarrow H(X^{213})\qquad ((3,2,1),(3,1,2))\nn\\
&\qquad \Leftrightarrow I(K_1;W_3Z_3|X^{213})\geq 0\nn\\
&H(W_2W_3Z_3K_1K_2)-H(W_1W_2Z_1)-H(W_3Z_3K_1)+H(W_1Z_1)\leq 0\\
&\qquad H(W_1W_2Z_1)\rightarrow H(W_2W_3Z_3)=H(W_2W_3Z_3K_2)\qquad ((2,3,1),(3,1,2))\nn\\
&\qquad H(W_1Z_1)\rightarrow H(W_3Z_3)\qquad ((3,1,2),(3,1,2))\nn\\
&\qquad \Leftrightarrow I(W_2K_2;K_1|W_3Z_3)\geq 0\nn\\
&3H(W_1W_2Z_1Z_2)-3H(W_1W_2Z_1Z_2X^{123})\leq 0\\
&\qquad \Leftrightarrow H(X^{123}|W_1W_2Z_1Z_2)\geq 0\nn\\
&3H(W_1W_2Z_1X^{123})-3H(W_1W_2Z_1Z_2X^{123})\leq 0\label{eq:conv1_last}\\
&\qquad \Leftrightarrow H(Z_2|W_1W_2Z_1X^{123})\geq 0\nn
\end{align}
by combining \eqref{eq:conv1_beg} to \eqref{eq:conv1_last}, we have:
\begin{equation}
4H(W_1W_2W_3)+H(W_1W_2)+H(W_1W_2Z_1X^{123})-H(W_1Z_1X^{123})-10H(Z_1)-6H(X^{123})+H(K_1)\leq 0\label{eq:conv1-mid}
\end{equation}
Note that
\begin{align}
H(K_1)&=I(Z_1X^{213};W_1)\\
&=H(Z_1X^{213})+H(W_1)-H(W_1Z_1X^{213})\\
&=H(W_2Z_1X^{213})+H(W_1)-H(W_1W_2Z_1X^{213})
\end{align}
and
\begin{align}
H(W_1Z_1X^{123}) \rightarrow H(W_2Z_1X^{213})\qquad ((2,1,3),(1,2,3))\\
H(W_1W_2Z_1X^{123}) \rightarrow H(W_1W_2Z_1X^{213})\qquad ((2,1,3),(1,2,3))\label{eq:conv1-final}
\end{align}
Thus, by combining \eqref{eq:conv1-mid} to \eqref{eq:conv1-final}, we conclude that 
\begin{align}
10M+6R\geq 15.
\end{align}

Similarly, for the lower bound $5M+4R\geq 9$, we have
\begin{align}
&H(W_1W_2W_3)-H(W_1W_2)-H(W_1W_3K_1K_2)+H(W_1K_1K_2)\leq 0\label{eq:conv2_beg}\\
&\qquad H(W_1W_2)=H(W_1W_2K_1K_2)\nn\\
&\qquad \Leftrightarrow I(W_2;W_3|W_1K_1K_2)\geq 0\nn\\
&2H(W_1W_2W_3)-2H(W_1W_2Z_1Z_2X^{123})-2H(W_1W_2Z_1X^{123}X^{231})+2H(W_1W_2Z_1X^{123})\leq 0\\
&\qquad H(W_1W_2W_3)=H(W_1W_2Z_1Z_2X^{123}X^{213})\nn\\
&\qquad \Leftrightarrow I(Z_2;X^{231}|W_1W_2Z_1X^{123})\geq 0\nn\\
&H(W_1W_2Z_1Z_2X^{123})-H(W_3Z_1Z_3K_1)-H(W_3Z_3X^{213}K_1)+H(W_3Z_3K_1)\leq 0\\
&\qquad H(W_1W_2Z_1Z_2X^{123})\rightarrow H(W_2W_3Z_1Z_3X^{213}) =H(W_3Z_1Z_3X^{213}K_1)\qquad ((3,2,1),(3,1,2))\nn\\
&\qquad \Leftrightarrow I(Z_1;X^{213}|W_3Z_3K_1)\geq 0\nn\\
&H(W_1W_2Z_1Z_2X^{123})-2H(W_1Z_1X^{123})+H(X^{123}K_2)\leq 0\\
&\qquad H(W_1W_2Z_1Z_2X^{123})=H(W_1W_2Z_1Z_2X^{123}K_2)\nn\\
&\qquad H(W_1Z_1X^{123})=H(W_1Z_1X^{123}K_2)\nn\\
&\qquad H(W_1Z_1X^{123})\rightarrow H(W_2Z_2X^{123})=H(W_2Z_2X^{123}K_2)\qquad ((2,3,1),(2,3,1))\nn\\
&\qquad \Leftrightarrow I(W_1Z_1;W_2Z_2|X^{123}K_2)\geq 0\nn\\
&H(W_1W_2Z_1X^{123}X^{231})-H(W_1W_2Z_1X^{123})-H(W_3Z_3X^{123}K_2)+H(W_3Z_3K_2)\leq 0\\
&\qquad H(W_1W_2Z_1X^{123}X^{231})\rightarrow H(W_2W_3Z_3X^{123}X^{312})=H(W_2W_3Z_3X^{123}X^{312}K_2)\qquad ((2,3,1),(3,1,2))\nn\\
&\qquad H(W_1W_2Z_1X^{123})\rightarrow H(W_2W_3Z_3X^{312})=H(W_2W_3Z_3X^{312}K_2)\qquad ((2,3,1),(3,1,2))\nn\\
&\qquad \Leftrightarrow I(W_2X^{312};X^{123}|W_3Z_3K_2)\geq 0\nn\\
&H(W_1W_2Z_1X^{123}X^{231})-H(W_1W_2Z_1X^{123})-H(W_1Z_1X^{123})+H(W_1Z_1K_1K_2)\leq 0\\
&\qquad H(W_1W_2Z_1X^{123}X^{231})=H(W_1W_2Z_1X^{123}X^{231}K_1K_2)\nn\\
&\qquad H(W_1W_2Z_1X^{123})\rightarrow H(W_1W_2Z_1X^{231})=H(W_1W_2Z_1X^{231}K_1K_2)\qquad ((2,1,3),(1,3,2))\nn\\
&\qquad H(W_1Z_1X^{123})=H(W_1Z_1X^{123}K_1K_2)\nn\\
&\qquad \Leftrightarrow I(W_2X^{231};X^{123}|W_1Z_1K_1K_2)\geq 0\nn\\
&H(W_1W_2Z_1X^{123})-H(W_1Z_1X^{123})-H(W_1Z_1K_1K_2)+H(Z_1K_1K_2)\leq 0\\
&\qquad H(W_1W_2Z_1X^{123})\rightarrow H(W_1W_2Z_1X^{213})=H(W_1W_2Z_1X^{213}K_1K_2)\qquad ((2,1,3),(1,2,3))\nn\\
&\qquad H(W_1Z_1X^{123})\rightarrow H(W_2Z_1X^{213})=H(W_2Z_1X^{213}K_1K_2)\qquad ((2,1,3),(1,2,3))\nn\\
&\qquad \Leftrightarrow I(W_1;W_2X^{213}|Z_1K_1K_2)\geq 0\nn\\
&H(W_1W_2Z_1X^{123})-H(W_1Z_1X^{123})-H(W_1X^{123})+H(X^{213}K_1)\leq 0\\
&\qquad H(W_1W_2Z_1X^{123})\rightarrow H(W_1W_2Z_1X^{231})=H(W_1W_2Z_1X^{231}K_1)\qquad ((2,1,3),(1,2,3))\nn\\
&\qquad H(W_1Z_1X^{123})\rightarrow H(W_2Z_1X^{231})=H(W_2Z_1X^{231}K_1)\qquad ((2,1,3),(1,2,3))\nn\\
&\qquad H(W_1X^{123})\rightarrow H(W_1X^{213})=H(W_1X^{213}K_1)\qquad ((1,3,2),(2,3,1))\nn\\
&\qquad \Leftrightarrow I(W_1;W_2Z_1|X^{213}K_1)\geq 0\nn\\
&H(W_1W_2)-H(W_1)-H(W_2K_1K_2)+H(K_1)\leq 0\\
&\qquad H(W_1W_2)=H(W_1W_2K_1K_2)\nn\\
&\qquad H(W_1)=H(W_1K_1)\nn\\
&\qquad \Leftrightarrow I(W_1;W_2K_2|K_1)\geq 0\nn\\
&5H(W_1Z_1X^{123})-5H(Z_1)-5H(X^{123})\leq 0\\
&\qquad H(W_1Z_1X^{123})=H(Z_1X^{123})\nn\\
&\qquad \Leftrightarrow I(Z_1;X^{123})\geq 0\nn\\
&H(W_3Z_3K_1K_2)-H(W_3Z_3K_1)-H(W_3Z_3K_2)+H(W_1Z_1)\leq 0\\
&\qquad H(W_1Z_1)\rightarrow H(W_3Z_3)\qquad ((3,1,2),(3,1,2))\nn\\
&\qquad \Leftrightarrow I(K_1;K_2|W_3Z_3)\geq 0\nn\\
&H(W_1W_3K_1K_2)-H(W_1K_1K_2)-H(W_3K_1K_2)+H(K_1K_2)\leq 0\\
&\qquad \Leftrightarrow I(W_1;W_3|K_1K_2)\geq 0\nn\\
&H(W_3Z_1Z_3K_1K_2)-H(W_3Z_1K_1K_2)-H(W_3Z_3K_1K_2)+H(W_3K_1K_2)\leq 0\\
&\qquad \Leftrightarrow I(Z_1;Z_3|W_3K_1K_2)\geq 0\nn\\
&H(W_3Z_1K_1K_2)-H(W_3Z_1K_2)-H(Z_1K_1K_2)+H(Z_1K_2)\leq 0\\
&\qquad \Leftrightarrow I(K_1;W_3|Z_1K_2)\geq 0\nn\\
&H(W_3Z_3X^{123}K_2)-H(Z_3K_2)-H(X^{123}K_2)+H(K_2)\leq 0\\
&\qquad H(W_3Z_3X^{123}K_2)=H(Z_3X^{123}K_2)\nn\\
&\qquad \Leftrightarrow I(Z_3;X^{123}|K_2)\geq 0\nn\\
&H(W_3Z_1Z_3K_1)-H(W_3Z_1Z_3K_1K_2)\leq 0\\
&\qquad \Leftrightarrow H(K_2|W_3Z_1Z_3K_1)\geq 0\nn\\
&H(Z_3K_2)-H(Z_1)-H(K_2)\leq 0\\
&\qquad H(Z_1)\rightarrow H(Z_3)\qquad ((3,1,2),(1,2,3))\nn\\
&\qquad \Leftrightarrow I(Z_3;K_2)\geq 0\nn\\
&H(W_2K_1K_2)-H(W_1)-H(K_1K_2)+H(K_2)\leq 0\\
&\qquad H(W_1)\rightarrow H(W_2)=H(W_2K_2)\qquad ((2,1,3),(1,2,3))\nn\\
&\qquad \Leftrightarrow I(W_2;K_1|K_2)\geq 0\nn\\
&H(W_3Z_1K_2)-H(W_1Z_1)-H(Z_1K_2)+H(Z_1)\leq 0\\
&\qquad H(W_1Z_1)\rightarrow H(W_3Z_1)\qquad ((3,1,2),(1,2,3))\nn\\
&\qquad \Leftrightarrow I(W_3;K_2|Z_1)\geq 0\nn\\
&H(W_3Z_3X^{213}K_1)-H(W_1Z_1X^{123})-H(X^{213}K_1)+H(X^{123})\leq 0\\
&\qquad H(W_1Z_1X^{123})\rightarrow H(W_3Z_3X^{213})\qquad ((3,2,1),(3,1,2))\nn\\
&\qquad H(X^{123})\rightarrow H(X^{213})\qquad ((3,2,1),(3,1,2))\nn\\
&\qquad \Leftrightarrow I(K_1;W_3Z_3|X^{213})\geq 0\nn\\
&H(W_1W_2X^{123})-H(W_1W_2Z_1X^{123})\leq 0\label{eq:conv2_last}\\
&\qquad \Leftrightarrow H(Z_1|W_1W_2X^{123})\geq 0\nn
\end{align}
by combining \eqref{eq:conv2_beg} to \eqref{eq:conv2_last}, we have:
\begin{align}
&3H(W_1W_2W_3)+H(W_1W_2Z_1X^{123})+H(W_1W_2X^{123})-H(W_1Z_1X^{123})\nn\\
&\qquad\qquad-H(W_1X^{123})-2H(W_1)-5H(Z_1)-4H(X^{123})+H(K_1)+H(K_2)\leq 0\label{eq:conv2-mid}
\end{align}
Note that
\begin{align}
H(K_1)&=I(Z_1X^{213};W_1)\\
&=H(Z_1X^{213})+H(W_1)-H(W_1Z_1X^{213})\\
&=H(W_2Z_1X^{213})+H(W_1)-H(W_1W_2Z_1X^{213})\\
H(K_2)&=I(W_1X^{123};W_2)\\
&=H(W_1X^{123})+H(W_2)-H(W_1W_2X^{123})
\end{align}
and
\begin{align}
H(W_1Z_1X^{123}) \rightarrow H(W_2Z_1X^{213})\qquad ((2,1,3),(1,2,3))\\
H(W_1W_2Z_1X^{123}) \rightarrow H(W_1W_2Z_1X^{213})\qquad ((2,1,3),(1,2,3))\label{eq:conv2-final}
\end{align}
Thus, by combining \eqref{eq:conv2-mid} to \eqref{eq:conv2-final}, we conclude that 
\begin{align}
5M+4R\geq 9
\end{align}

\subsection{The transform mappings in Lemma \ref{lem:sig}}\label{apx:sig-map}
The transform mapping for Part 2 of Table \ref{tab:sig-part} is provided as follow:
\begin{equation}
\mathbf{X^{ABC}} \xrightarrow{\quad f \quad} \mathbf{X^{BCA}} \xrightarrow{\quad f \quad} \mathbf{X^{CAB}} \xrightarrow{\quad f \quad} \mathbf{X^{ABC}}
\end{equation}

The transform mapping for Part 3 of Table \ref{tab:sig-part} is provided as follow:
\begin{equation}
\mathbf{X^{ACB}} \xrightarrow{\quad f^2 \quad} \mathbf{X^{CBA}} \xrightarrow{\quad f^2 \quad} \mathbf{X^{BAC}} \xrightarrow{\quad f^2 \quad} \mathbf{X^{ACB}}
\end{equation}

The transform mapping for Part 4 of Table \ref{tab:sig-part} is provided as follow:
\begin{equation}
\begin{aligned}
&\mathbf{X^{ABB}} &\xrightarrow{\quad g \quad}\quad &\mathbf{X^{BAB}} &\xrightarrow{\quad g \quad}\quad &\mathbf{X^{BBA}} &\xrightarrow{\quad g \quad}\quad &\mathbf{X^{ABB}}\\
&\Bigg\downarrow f& & \Bigg\downarrow f & &\Bigg\downarrow f & &\Bigg\downarrow f\\
&\mathbf{X^{BCC}} &\xrightarrow{\quad g \quad}\quad &\mathbf{X^{CBC}} &\xrightarrow{\quad g \quad}\quad &\mathbf{X^{CCB}} &\xrightarrow{\quad g \quad}\quad &\mathbf{X^{BCC}}\\
&\Bigg\downarrow f & &\Bigg\downarrow f & &\Bigg\downarrow f & &\Bigg\downarrow f\\
&\mathbf{X^{CAA}} &\xrightarrow{\quad g \quad}\quad &\mathbf{X^{ACA}} &\xrightarrow{\quad g \quad}\quad &\mathbf{X^{AAC}} &\xrightarrow{\quad g \quad}\quad &\mathbf{X^{CAA}}\\
&\Bigg\downarrow f& & \Bigg\downarrow f & &\Bigg\downarrow f & &\Bigg\downarrow f\\
&\mathbf{X^{ABB}} &\xrightarrow{\quad g \quad}\quad &\mathbf{X^{BAB}} &\xrightarrow{\quad g \quad}\quad &\mathbf{X^{BBA}} &\xrightarrow{\quad g \quad}\quad &\mathbf{X^{ABB}}
\end{aligned}
\end{equation}

The transform mapping for Part 5 of Table \ref{tab:sig-part} is provided as follow:
\begin{equation}
\begin{aligned}
&\mathbf{X^{ACC}} &\xrightarrow{\quad g \quad}\quad &\mathbf{X^{CAC}} &\xrightarrow{\quad g \quad} \quad &\mathbf{X^{CCA}} &\xrightarrow{\quad g \quad}\quad &\mathbf{X^{ACC}}\\
&\Bigg\downarrow f^2& & \Bigg\downarrow f^2 & &\Bigg\downarrow f^2 & &\Bigg\downarrow f^2\\
&\mathbf{X^{CBB}} &\xrightarrow{\quad g \quad}\quad &\mathbf{X^{BCB}} &\xrightarrow{\quad g \quad}\quad &\mathbf{X^{BBC}} &\xrightarrow{\quad g \quad}\quad &\mathbf{X^{CBB}}\\
&\Bigg\downarrow f^2& & \Bigg\downarrow f^2 & &\Bigg\downarrow f^2 & &\Bigg\downarrow f^2\\
&\mathbf{X^{BAA}} &\xrightarrow{\quad g \quad}\quad &\mathbf{X^{ABA}} &\xrightarrow{\quad g \quad}\quad &\mathbf{X^{AAB}} &\xrightarrow{\quad g \quad}\quad &\mathbf{X^{BAA}}\\
&\Bigg\downarrow f^2& & \Bigg\downarrow f^2 & &\Bigg\downarrow f^2 & &\Bigg\downarrow f^2\\
&\mathbf{X^{ACC}} &\xrightarrow{\quad g \quad}\quad &\mathbf{X^{CAC}} &\xrightarrow{\quad g \quad} \quad &\mathbf{X^{CCA}} &\xrightarrow{\quad g \quad}\quad &\mathbf{X^{ACC}}
\end{aligned}
\end{equation}

\newpage
\subsection{The decoding processes for the corner point $(0.6,1.5)$}\label{apx:dec}
Restate the linear scheme as follow:
\begin{table}[h!]
\caption{The linear scheme for the corner point $(0.6,1.5)$.}
\begin{center}
\begin{tabular}{|c|c|c|c|}
    \hline
    \multirow{2}{*}{$Z_1$}& $A_1\oplus A_3\oplus A_4\oplus B_2\oplus B_{10}$ & $B_1\oplus B_3\oplus B_4\oplus C_2\oplus C_{10}$ & $C_1\oplus C_3\oplus C_4\oplus A_2\oplus A_{10}$\\
    \cline{2-4}
    & $A_2\oplus B_1$ & $B_2\oplus C_1$ & $C_2\oplus A_1$\\
    \hline
    \multirow{2}{*}{$Z_2$}& $A_4\oplus A_6\oplus A_7\oplus B_5\oplus B_{10}$ & $B_4\oplus B_6\oplus B_7\oplus C_5\oplus C_{10}$ & $C_4\oplus C_6\oplus C_7\oplus A_5\oplus A_{10}$\\
    \cline{2-4}
    & $A_5\oplus B_4$ & $B_5\oplus C_4$ & $C_5\oplus A_4$\\
    \hline
    \multirow{2}{*}{$Z_3$}& $A_7\oplus A_9\oplus A_1\oplus B_8\oplus B_{10}$ & $B_7\oplus B_9\oplus B_1\oplus C_8\oplus C_{10}$ & $C_7\oplus C_9\oplus C_1\oplus A_8\oplus A_{10}$\\
    \cline{2-4}
    & $A_8\oplus B_7$ & $B_8\oplus C_7$ & $C_8\oplus A_7$\\
    \hline
     \multirow{5}{*}{$X^{ABC}$} & $B_2\oplus B_{10}$ & $C_5\oplus C_{10}$ & $A_8\oplus A_{10}$\\
    \cline{2-4}
     &$A_4\oplus A_6\oplus A_7$ & $B_7\oplus B_9\oplus B_1$ & $C_1\oplus C_3\oplus C_4$\\
    \cline{2-4}
     &$A_7$ & $B_1$ & $C_4$\\
    \cline{2-4}
     &$A_5$ & $B_8$ & $C_2$\\
    \cline{2-4}
      & $A_1\oplus A_9\oplus B_2\oplus C_1$ & $B_3\oplus B_4\oplus C_5\oplus A_4$ & $C_6\oplus C_7\oplus A_8\oplus B_7$\\
    \hline
     \multirow{5}{*}{$X^{ACB}$} &$A_5\oplus A_{10}$ & $B_2\oplus B_{10}$ & $C_8\oplus C_{10}$\\
    \cline{2-4}
     &$A_7\oplus A_9\oplus A_1$ & $B_4\oplus B_6\oplus B_7$ & $C_1\oplus C_3\oplus C_4$\\
    \cline{2-4}
     &$A_4$ & $B_1$ & $C_7$\\
    \cline{2-4}
     &$A_8$ & $B_5$ & $C_2$\\
    \cline{2-4}
     &$A_6\oplus A_7\oplus B_2\oplus C_1$ & $B_3\oplus B_4\oplus C_8\oplus A_7$ & $C_9\oplus C_1\oplus A_5\oplus B_4$\\
    \hline
     \multirow{5}{*}{$X^{ABB}$}& $B_2\oplus B_{10}$ & $A_4\oplus A_6\oplus A_7$ & $A_7\oplus A_9\oplus A_1$\\
    \cline{2-4}
     &$B_3\oplus B_4$ & $A_5\oplus C_{10}\oplus A_4$ & $A_8\oplus C_{10}\oplus A_7$\\
    \cline{2-4}
     &$A_{10}$ & $B_{10}$ & $C_{10}$\\
    \cline{2-4}
     &$B_1$ & $A_5$ & $A_8$\\
    \cline{2-4}
    &$B_4\oplus B_7$ & $B_6\oplus B_9$ & $B_5\oplus B_8$\\
    \hline
     \multirow{5}{*}{$X^{ACC}$}& $C_1\oplus C_3\oplus C_4$ & $A_5\oplus A_{10}$ & $A_8\oplus A_{10}$\\
    \cline{2-4}
     &$C_1\oplus C_2$ & $A_6\oplus A_7$ & $A_9\oplus A_1$\\
    \cline{2-4}
     &$A_{10}$ & $B_{10}$ & $C_{10}$\\
    \cline{2-4}
     &$C_2$ & $A_4$ & $A_7$\\
    \cline{2-4}
     &$C_5\oplus C_8$ & $C_4\oplus C_7$ & $C_6\oplus C_9$\\
    \hline
\end{tabular}
\end{center}
\end{table}

The decoding process is given in Table \ref{tab:dec}, where $[\cdot]$ means the content is in the caches, $(\cdot)$ means the content is from the delivery messages, and $\{\cdot\}$ means the content is the previous decoding result.

\begin{longtable}[ht]{|c|c|c|c|}
\caption{The decoding process for the corner point $(0.6,1.5)$.}\label{tab:dec}\\

\hline \multicolumn{1}{|c|}{\textbf{Demand}} & \multicolumn{1}{c|}{\textbf{User}} &
\multicolumn{1}{c|}{\textbf{decoding output}}&
\multicolumn{1}{c|}{\textbf{decoding input}} \\ \hline 
\endfirsthead

\multicolumn{4}{c}%
{{\bfseries \tablename\ \thetable{} -- continued from previous page}} \\
\hline \multicolumn{1}{|c|}{\textbf{Demand}} & \multicolumn{1}{c|}{\textbf{User}} &
\multicolumn{1}{c|}{\textbf{decoding output}}&
\multicolumn{1}{c|}{\textbf{decoding input}} \\ \hline 
\endhead

\hline \multicolumn{4}{|r|}{{Continued on next page}} \\ \hline
\endfoot

\hline \hline
\endlastfoot

\hline
\multirow{11}{*}{$ABC$}&\multirow{11}{*}{$Z_1$}& $A_1\oplus A_3\oplus A_4$ & $[A_1\oplus A_3\oplus A_4\oplus B_2\oplus B_{10}]\oplus (B_2\oplus B_{10})$\\
\cline{3-4}
& & $A_2\oplus A_{10}$ & $[C_1\oplus C_3\oplus C_4\oplus A_2\oplus A_{10}]\oplus (C_1\oplus C_3\oplus C_4)$\\
\cline{3-4}
& & $A_1$ & $[C_2\oplus A_1]\oplus (C_2)$\\
\cline{3-4}
& & $A_2$ & $[A_2\oplus B_1]\oplus (B_1)$\\
\cline{3-4}
& & $A_{10}$ & $\{A_2\oplus A_{10}\}\oplus \{A_2\}$\\
\cline{3-4}
& & $A_9$ & $[B_2\oplus C_1]\oplus (A_1\oplus A_9\oplus B_2\oplus C_1)\oplus \{A_1\}$\\
\cline{3-4}
& & \multirow{2}{*}{$A_4$} & $[B_1\oplus B_3\oplus B_4\oplus C_2\oplus C_{10}]\oplus (B_1)\oplus (C_2)$\\
& & & $\oplus (C_5\oplus C_{10})\oplus (B_3\oplus B_4\oplus C_5\oplus A_4)$\\
\cline{3-4}
& & $A_3$ & $\{A_1\oplus A_3\oplus A_4\}\oplus \{A_1\}\oplus \{A_4\}$\\
\cline{3-4}
& & $A_6$ & $(A_4\oplus A_6\oplus A_7)\oplus \{A_4\}\oplus (A_7)$\\
\cline{3-4}
& & $A_{8}$ & $(A_8\oplus A_{10})\oplus \{A_{10}\}$\\
\cline{2-4}
\multirow{22}{*}{$ABC$} & \multirow{11}{*}{$Z_2$} & $B_4\oplus B_6\oplus B_7$ & $[B_4\oplus B_6\oplus B_7\oplus C_5\oplus C_{10}]\oplus (C_5\oplus C_{10})$\\
\cline{3-4}
& & $B_5\oplus B_{10}$ & $[A_4\oplus A_6\oplus A_7\oplus B_5\oplus A_{10}]\oplus (A_4\oplus A_6\oplus A_7)$\\
\cline{3-4}
& & $B_4$ & $[A_5\oplus B_4]\oplus (A_5)$\\
\cline{3-4}
& & $B_5$ & $[B_5\oplus C_4]\oplus (C_4)$\\
\cline{3-4}
 & & $B_{10}$ & $\{B_5\oplus B_{10}\}\oplus \{B_5\}$\\
\cline{3-4}
& & $B_3$ & $[C_5\oplus A_4]\oplus (B_3\oplus B_4\oplus C_5\oplus A_4)\oplus \{B_4\}$\\
\cline{3-4}
& & \multirow{2}{*}{$B_7$} & $[C_4\oplus C_6\oplus C_7\oplus A_5\oplus A_{10}]\oplus (C_4)\oplus (A_5)$\\
& & & $\oplus (A_8\oplus A_{10})\oplus (C_6\oplus C_7\oplus A_8\oplus B_7)$\\
\cline{3-4}
& & $B_6$ & $\{B_4\oplus B_6\oplus B_7\}\oplus \{B_4\}\oplus \{B_7\}$\\
\cline{3-4}
& & $B_9$ & $(B_7\oplus B_9\oplus B_1)\oplus \{B_7\}\oplus (B_1)$\\
\cline{3-4}
& & $B_{2}$ & $(B_2\oplus B_{10})\oplus \{B_{10}\}$\\
\cline{2-4}
& \multirow{11}{*}{$Z_3$} & $C_7\oplus C_9\oplus C_1$ & $[C_7\oplus C_9\oplus C_1\oplus A_8\oplus A_{10}]\oplus (A_8\oplus A_{10})$\\
\cline{3-4}
& & $C_8\oplus C_{10}$ & $[B_7\oplus B_9\oplus B_1\oplus C_8\oplus C_{10}]\oplus (B_7\oplus B_9\oplus B_1)$\\
\cline{3-4}
& & $C_7$ & $[B_8\oplus C_7]\oplus (B_8)$\\
\cline{3-4}
& & $C_8$ & $[C_8\oplus A_7]\oplus (A_7)$\\
\cline{3-4}
& & $C_{10}$ & $\{C_8\oplus C_{10}\}\oplus \{C_8\}$\\
\cline{3-4}
& & $C_6$ & $[A_8\oplus B_7]\oplus (C_6\oplus C_7\oplus A_8\oplus B_7)\oplus \{C_7\}$\\
\cline{3-4}
& & \multirow{2}{*}{$C_1$} & $[A_7\oplus A_9\oplus A_1\oplus B_8\oplus B_{10}]\oplus (A_7)\oplus (B_8)$\\
& & & $\oplus (B_2\oplus B_{10})\oplus (A_1\oplus A_9\oplus B_2\oplus C_1)$\\
\cline{3-4}
& & $C_9$ & $\{C_7\oplus C_9\oplus C_1\}\oplus \{C_7\}\oplus \{C_1\}$\\
\cline{3-4}
& & $C_3$ & $(C_1\oplus C_3\oplus C_4)\oplus \{C_1\}\oplus (C_4)$\\
\cline{3-4}
& & $C_{5}$ & $(C_5\oplus C_{10})\oplus \{C_{10}\}$\\
\hline
\multirow{19}{*}{$ACB$} & \multirow{11}{*}{$Z_1$} & $A_1\oplus A_3\oplus A_4$ & $[A_1\oplus A_3\oplus A_4\oplus B_2\oplus B_{10}]\oplus (B_2\oplus B_{10})$\\
\cline{3-4}
& & $A_2\oplus A_{10}$ & $[C_1\oplus C_3\oplus C_4\oplus A_2\oplus A_{10}]\oplus (C_1\oplus C_3\oplus C_4)$\\
\cline{3-4}
& & $A_1$ & $[C_2\oplus A_1]\oplus (C_2)$\\
\cline{3-4}
& & $A_2$ & $[A_2\oplus B_1]\oplus (B_1)$\\
\cline{3-4}
& & $A_{10}$ & $\{A_2\oplus A_{10}\}\oplus \{A_2\}$\\
\cline{3-4}
& & \multirow{2}{*}{$A_7$} & $[B_1\oplus B_3\oplus B_4\oplus C_2\oplus C_{10}]\oplus (B_1)\oplus (C_2)$\\
& & & $\oplus (C_8\oplus C_{10})\oplus (B_3\oplus B_4\oplus C_8\oplus A_7)$\\
\cline{3-4}
& & $A_6$ & $[B_2\oplus C_1]\oplus (A_6\oplus A_7\oplus B_2\oplus C_1)\oplus \{A_7\}$\\
\cline{3-4}
& & $A_3$ & $\{A_1\oplus A_3\oplus A_4\}\oplus \{A_1\}\oplus \{A_4\}$\\
\cline{3-4}
& & $A_9$ & $(A_7\oplus A_9\oplus A_1)\oplus \{A_1\}\oplus (A_7)$\\
\cline{3-4}
& & $A_{5}$ & $(A_5\oplus A_{10})\oplus \{A_{10}\}$\\
\cline{2-4}
&\multirow{8}{*}{$Z_2$} & $C_4\oplus C_6\oplus C_7$ & $[C_4\oplus C_6\oplus C_7\oplus A_5\oplus A_{10}]\oplus (A_5\oplus B_{10})$\\
\cline{3-4}
& & $C_5\oplus C_{10}$ & $[B_4\oplus B_6\oplus B_7\oplus C_5\oplus A_{10}]\oplus (B_4\oplus B_6\oplus B_7)$\\
\cline{3-4}
& & $C_4$ & $[B_5\oplus C_4]\oplus (B_5)$\\
\cline{3-4}
& & $C_5$ & $[C_5\oplus A_4]\oplus (A_4)$\\
\cline{3-4}
& & $C_{10}$ & $\{C_5\oplus C_{10}\}\oplus \{C_5\}$\\
\cline{3-4}
&  & \multirow{2}{*}{$C_1$} & $[A_4\oplus A_6\oplus A_7\oplus B_5\oplus B_{10}]\oplus (A_4)\oplus (B_5)$\\
& & & $\oplus (B_2\oplus B_{10})\oplus (A_6\oplus A_7\oplus B_2\oplus C_1)$\\
\cline{3-4}
& & $C_9$ & $[A_5\oplus B_4]\oplus (C_9\oplus C_1\oplus A_5\oplus B_4)\oplus \{C_1\}$\\
\cline{3-4}
\multirow{14}{*}{$ACB$} & \multirow{3}{*}{$Z_2$} & $C_6$ & $\{C_4\oplus C_6\oplus C_7\}\oplus \{C_4\}\oplus \{C_7\}$\\
\cline{3-4}
& & $C_3$ & $(C_1\oplus C_3\oplus C_4)\oplus \{C_4\}\oplus (C_1)$\\
\cline{3-4}
& & $C_{8}$ & $(C_8\oplus C_{10})\oplus \{C_{10}\}$\\
\cline{2-4}
& \multirow{11}{*}{$Z_3$} & $B_7\oplus B_9\oplus B_1$ & $[B_7\oplus B_9\oplus B_1\oplus C_8\oplus A_{10}]\oplus (C_8\oplus B_{10})$\\
\cline{3-4}
& & $B_8\oplus B_{10}$ & $[A_7\oplus A_9\oplus A_1\oplus B_8\oplus B_{10}]\oplus (A_7\oplus A_9\oplus A_1)$\\
\cline{3-4}
& & $B_7$ & $[A_8\oplus B_7]\oplus (A_8)$\\
\cline{3-4}
 &  & $B_8$ & $[B_8\oplus C_7]\oplus (C_7)$\\
\cline{3-4}
& & $B_{10}$ & $\{B_8\oplus B_{10}\}\oplus \{B_8\}$\\
\cline{3-4}
&  & \multirow{2}{*}{$B_4$} & $[C_7\oplus C_9\oplus C_1\oplus A_8\oplus A_{10}]\oplus (C_7)\oplus (A_8)$\\
& & & $\oplus (A_5\oplus A_{10})\oplus (C_9\oplus C_1\oplus A_5\oplus B_4)$\\
\cline{3-4}
& & $B_3$ & $[C_8\oplus A_7]\oplus (B_3\oplus B_4\oplus C_8\oplus A_7)\oplus \{B_4\}$\\
\cline{3-4}
& & $B_9$ & $\{B_7\oplus B_9\oplus B_1\}\oplus \{B_7\}\oplus \{B_1\}$\\
\cline{3-4}
& & $B_6$ & $(B_4\oplus B_6\oplus B_7)\oplus \{B_7\}\oplus (B_4)$\\
\cline{3-4}
& & $B_{2}$ & $(B_2\oplus B_{10})\oplus \{B_{10}\}$\\
\hline
	\multirow{27}{*}{$ABB$} & \multirow{13}{*}{$Z_1$} & $A_1\oplus A_3\oplus A_4$ & $[A_1\oplus A_3\oplus A_4\oplus B_2\oplus B_{10}]\oplus (B_2\oplus B_{10})$ \\
	\cline{3-4}
	 &  & \multirow{2}{*}{$A_2\oplus C_{10}\oplus A_1$}  & $[A_2\oplus B_1]\oplus [B_1\oplus B_3\oplus B_4\oplus C_2\oplus C_{10}]$\\
	 & & & $\oplus [C_2\oplus A_1]\oplus (B_3\oplus B_4)$\\
	\cline{3-4}
	 & & $A_1\oplus A_2$ &  $\{A_2\oplus C_{10}\oplus A_1\}\oplus (C_{10})$\\
	 \cline{3-4}
	 & & $A_4\oplus A_5$ &  $(A_5\oplus C_{10}\oplus A_4)\oplus (C_{10})$\\
	 \cline{3-4}
	 & & $A_7\oplus A_8$ &  $(A_8\oplus C_{10}\oplus A_7)\oplus (C_{10})$\\
    \cline{3-4}
    &  & $A_2$ & $[A_2\oplus B_1]\oplus (B_1)$\\
    \cline{3-4}
    & & $A_4$ & $\{A_4\oplus A_5\}\oplus (A_5)$\\
    \cline{3-4}
    & & $A_7$ & $\{A_7\oplus A_8\}\oplus (A_8)$\\
    \cline{3-4}
    &   & $A_1$ & $\{A_1\oplus A_2\}\oplus \{A_2\}$\\
    \cline{3-4}
    & & $A_3$ & $\{A_1\oplus A_3\oplus A_4\}\oplus \{A_1\}\oplus \{A_4\}$\\
    \cline{3-4}
    & & $A_6$ & $(A_4\oplus A_6\oplus A_7)\oplus \{A_4\}\oplus \{A_7\}$\\
    \cline{3-4}
    & & $A_9$ & $(A_7\oplus A_9\oplus A_1)\oplus \{A_7\}\oplus \{A_1\}$\\
    \cline{2-4}
    & \multirow{11}{*}{$Z_2$}  & $B_5\oplus B_{10}$ & $[A_4\oplus A_6\oplus A_7\oplus B_5\oplus B_{10}]\oplus (A_4\oplus A_6\oplus A_7)$ \\
	\cline{3-4}
    &  & \multirow{2}{*}{$B_6\oplus B_7$} & $[A_5\oplus B_4]\oplus [B_4\oplus B_6\oplus B_7\oplus C_5\oplus C_{10}]$\\
    & & & $\oplus [C_5\oplus A_4]\oplus (A_5\oplus C_{10}\oplus A_4)$\\
    \cline{3-4}
    & & $B_2$ & $(B_2\oplus B_{10})\oplus (B_{10})$\\
    \cline{3-4}
    & & $B_5$ & $\{B_5\oplus B_{10}\}\oplus (B_{10})$\\
    \cline{3-4}
    &  & $B_4$ & $[A_5\oplus B_4]\oplus (A_5)$\\
    \cline{3-4}
    & & $B_3$ & $(B_3\oplus B_4)\oplus \{B_4\}$\\
    \cline{3-4}
    & & $B_7$ & $(B_4\oplus B_7)\oplus \{B_4\}$\\
    \cline{3-4}
    & & $B_8$ & $(B_5\oplus B_8)\oplus \{B_5\}$\\
    \cline{3-4}
    &  & $B_6$ & $\{B_6\oplus B_7\}\oplus \{B_7\}$\\
    \cline{3-4}
    & & $B_9$ & $(B_6\oplus B_9)\oplus \{B_6\}$\\
    \cline{2-4}
    & \multirow{3}{*}{$Z_3$}  & $B_8\oplus B_{10}$ & $[A_7\oplus A_9\oplus A_1\oplus B_8\oplus B_{10}]\oplus (A_7\oplus A_9\oplus A_1)$ \\
	\cline{3-4}
    &   & \multirow{2}{*}{$B_9\oplus B_1$} & $[A_8\oplus B_7]\oplus [B_7\oplus B_9\oplus B_1\oplus C_8\oplus C_{10}]$\\
    & & & $\oplus [C_8\oplus A_7]\oplus (A_8\oplus C_{10}\oplus A_7)$\\
    \cline{3-4}
    & \multirow{8}{*}{$Z_3$} & $B_2$ & $(B_2\oplus B_{10})\oplus (B_{10})$\\
    \cline{3-4}
    & & $B_8$ & $\{B_8\oplus B_{10}\}\oplus (B_{10})$\\
    \cline{3-4}
    &   & $B_7$ & $[A_8\oplus B_7]\oplus (A_8)$\\
    \cline{3-4}
    & & $B_9$ & $(B_9\oplus B_1)\oplus \{B_1\}$\\
    \cline{3-4}
    & & $B_4$ & $(B_4\oplus B_7)\oplus \{B_7\}$\\
    \cline{3-4}
    & & $B_5$ & $(B_5\oplus B_8)\oplus \{B_8\}$\\
    \cline{3-4}
    &   & $B_3$ & $(B_3\oplus B_4)\oplus \{B_4\}$\\
    \cline{3-4}
    & & $B_6$ & $(B_6\oplus B_9)\oplus \{B_9\}$\\
    \hline
	\multirow{33}{*}{$ACC$} & \multirow{11}{*}{$Z_1$} &  $A_2\oplus A_{10}$ & $[C_1\oplus C_3\oplus C_4\oplus A_2\oplus A_{10}]\oplus (C_1\oplus C_3\oplus C_4)$ \\
	\cline{3-4}
	 &  & \multirow{2}{*}{$A_3\oplus A_4\oplus B_{10}$}  & $[C_2\oplus A_1]\oplus [A_1\oplus A_3\oplus A_4\oplus B_2\oplus B_{10}]$\\
	 & & & $\oplus [B_2\oplus C_1]\oplus (C_1\oplus C_2)$\\
	\cline{3-4}
	 & & $A_3\oplus A_4$ &  $\{A_3\oplus A_4\oplus B_{10}\}\oplus (B_{10})$\\
	 \cline{3-4}
	 & & $A_5$ &  $(A_5\oplus A_{10})\oplus (A_{10})$\\
	 \cline{3-4}
	 & & $A_8$ &  $(A_8\oplus A_{10})\oplus (A_{10})$\\
	 \cline{3-4}
	 & & $A_2$ &  $\{A_2\oplus A_{10}\}\oplus (A_{10})$\\
    \cline{3-4}
    &   & $A_1$ & $[C_2\oplus A_1]\oplus (C_2)$\\
    \cline{3-4}
    & & $A_3$ & $\{A_3\oplus A_4\}\oplus (A_4)$\\
    \cline{3-4}
    & & $A_6$ & $(A_6\oplus A_7)\oplus (A_7)$\\
    \cline{3-4}
    &   & $A_9$ & $(A_1\oplus A_9)\oplus \{A_1\}$\\
    \cline{2-4}
    & \multirow{12}{*}{$Z_2$}  &   $C_4\oplus C_6\oplus C_7$ & $[C_4\oplus C_6\oplus C_7\oplus A_5\oplus A_{10}]\oplus (A_5\oplus A_{10})$ \\
	\cline{3-4}
    &   & \multirow{2}{*}{$C_5\oplus B_{10}\oplus C_4$} & $[C_5\oplus A_4]\oplus [A_4\oplus A_6\oplus A_7\oplus B_5\oplus B_{10}]$\\
    & & & $\oplus [B_5\oplus C_4]\oplus (A_6\oplus A_7)$\\
    \cline{3-4}
    & & $C_5\oplus C_4$ & $(C_5\oplus B_{10}\oplus C_4)\oplus (B_{10})$\\
    \cline{3-4}
    &   &$C_1$ & $(C_1\oplus C_2)\oplus (C_2)$\\
    \cline{3-4}
    & & $C_5$ & $[C_5\oplus A_4]\oplus (A_4)$\\
    \cline{3-4}
    & & $C_8$ & $(C_5\oplus C_8)\oplus \{C_5\}$\\
    \cline{3-4}
    & & $C_4$ & $\{C_5\oplus C_4\}\oplus \{C_4\}$\\
    \cline{3-4}
    & & $C_7$ & $(C_4\oplus C_7)\oplus \{C_4\}$\\
    \cline{3-4}
    & & $C_6$ & $\{C_4\oplus C_6\oplus C_7\}\oplus (C_4\oplus C_7)$\\
    \cline{3-4}
    & & $C_9$ & $(C_6\oplus C_9)\oplus \{C_6\}$\\
    \cline{3-4}
    &   & $C_3$ & $(C_1\oplus C_3\oplus C_4)\oplus \{C_1\}\oplus \{C_4\}$\\
    \cline{2-4}
    & \multirow{10}{*}{$Z_3$}  & $C_7\oplus C_9\oplus C_1$ & $[C_7\oplus C_9\oplus C_1\oplus A_8\oplus A_{10}]\oplus (A_5\oplus A_{10})$ \\
	\cline{3-4}
    &   & \multirow{2}{*}{$C_8\oplus B_{10}\oplus C_7$} & $[C_8\oplus A_7]\oplus [A_7\oplus A_9\oplus A_1\oplus B_8\oplus B_{10}]$\\
    & & & $\oplus [B_8\oplus C_7]\oplus (A_9\oplus A_1)$\\
    \cline{3-4}
    & & $C_8\oplus C_7$ & $(C_8\oplus B_{10}\oplus C_7)\oplus (B_{10})$\\
    \cline{3-4}
    &   &$C_1$ & $(C_1\oplus C_2)\oplus (C_2)$ \\
    \cline{3-4}
    & & $C_8$ & $[C_8\oplus A_7]\oplus (A_7)$\\
    \cline{3-4}
    & & $C_5$ & $(C_5\oplus C_8)\oplus \{C_8\}$\\
    \cline{3-4}
    & & $C_7$ & $\{C_8\oplus C_7\}\oplus \{C_8\}$\\
    \cline{3-4}
    & & $C_4$ & $(C_4\oplus C_7)\oplus \{C_7\}$\\
    \cline{3-4}
    & & $C_9$ & $\{C_7\oplus C_9\oplus C_1\}\oplus \{C_1\}\oplus \{C_7\}$\\
    \cline{3-4}
    \multirow{2}{*}{$ACC$} & \multirow{2}{*}{$Z_3$} & $C_6$ & $(C_6\oplus C_9)\oplus \{C_9\}$\\
    \cline{3-4}
    &   & $C_3$ & $(C_1\oplus C_3\oplus C_4)\oplus \{C_1\}\oplus \{C_4\}$\\
    \hline
\end{longtable}

\newpage
\subsection{A linear scheme of Point $(M,R)=(0.5,\frac{5}{3})$ with coded content }
We may follow the achievability proof of Theorem \ref{TH1} to construct a linear scheme of Point $(M,R)=(0.5,\frac{5}{3})$ with coded content. The cache construction is given as follow:

\begin{table}[ht]
\begin{center}
\begin{tabular}{|c|c|c|c|}
\hline
$Z_1$ & $A_1\oplus A_2\oplus A_3\oplus B_3$ & $B_1\oplus B_2\oplus B_3\oplus C_3$ & $C_1\oplus C_2\oplus C_3\oplus A_3$ \\
\hline
$Z_2$ & $A_3\oplus A_4\oplus A_5\oplus B_5$ & $B_3\oplus B_4\oplus B_5\oplus C_5$ & $C_3\oplus C_4\oplus C_5\oplus A_5$ \\
\hline
$Z_3$ & $A_5\oplus A_6\oplus A_1\oplus B_1$ & $B_5\oplus B_6\oplus B_1\oplus C_1$ & $C_5\oplus C_6\oplus C_1\oplus A_1$ \\
\hline
\end{tabular}
\end{center}
\end{table}

and the corresponding construction of delivery messages are provided in the following table. 

\begin{table}[ht]
\begin{center}
\begin{tabular}{|c|c|c|}
	\hline
	\multicolumn{3}{|c|}{AAA}\\
    \hline
    \multicolumn{3}{|c|}{$A$}\\
    \hline
	\multicolumn{3}{|c|}{ABC}\\
    \hline
     $C_1\oplus C_2\oplus C_3$ & $A_3\oplus A_4\oplus A_5$ & $B_5\oplus B_6\oplus B_1$\\
    \hline
     $B_3$ & $C_5$ & $A_1$\\
    \hline
     \multicolumn{3}{|c|}{$B_1\oplus B_2\oplus B_3\oplus C_3\oplus C_4\oplus C_5\oplus A_5\oplus A_6\oplus A_1$}\\
    \hline
     $C_4$ & $A_6$ & $B_2$\\
    \hline
    \multicolumn{3}{|c|}{ACB}\\
    \hline
     $C_1\oplus C_2\oplus C_3$ & $B_3\oplus B_4\oplus B_5$ & $A_5\oplus A_6\oplus A_1$\\
    \hline
     $B_3$ & $A_5$ & $C_1$\\
    \hline
     \multicolumn{3}{|c|}{$B_1\oplus B_2\oplus B_3\oplus A_3\oplus A_4\oplus A_5\oplus C_5\oplus C_6\oplus C_1$}\\
    \hline
     $C_3\oplus C_6$ & $B_2\oplus B_5$ & $A_1\oplus A_4$\\
    \hline
    \multicolumn{3}{|c|}{ABB}\\
    \hline
     $A_3\oplus A_4\oplus A_5$ & $A_5\oplus A_6\oplus A_1$ & $B_1\oplus B_5$\\
    \hline
     \multicolumn{3}{|c|}{$B_2$\qquad $B_3$\qquad $B_4$\qquad $B_6$}\\
    \hline
     $A_1$ & $A_3$ & $A_5$\\
    \hline
    \multicolumn{3}{|c|}{ACC}\\
    \hline
     $A_1$ & $A_5$ & $C_1\oplus C_2\oplus C_3$\\
    \hline
     $C_1$ & $C_5$ & $C_3$\\
    \hline
     \multicolumn{3}{|c|}{$C_4\oplus C_6$}\\
    \hline
     $A_1\oplus A_4\oplus A_5$ & $A_5\oplus A_6\oplus A_1$ & $B_3$\\
    \hline
\end{tabular}
\end{center}
\end{table}

\subsection{A slight improvement for the lower bound of $R^*(M)$}
\begin{figure}[ht]
	\center
	\begin{tikzpicture}
	\center
	\begin{axis}[
	scale=1.5,
    xlabel={Cache size $M$},
    ylabel={Rate $R$},
    xmin=1/3, xmax=1,
    ymin=1, ymax=2,
    xtick={0.333333333,0.5,0.66666666,1},
    xticklabels={1/3,1/2,2/3,1},
    ytick={1,1.33333333,1.6666666666,2},
    yticklabels={1,\Large $\frac{4}{3}$,\Large $\frac{5}{3}$,2},
    legend pos=north east,
    ymajorgrids=true,
    grid style=dashed,
    ]
    \addplot[
    name path=ach,
	very thick,
	dashdotted,]
    coordinates {
    (1/3,2)(1/2,5/3)(1,1)
    };


    \addplot[
    name path=ach,
    color=red,
    mark=o,
	thick,]
    coordinates {
    (1/3,2)(1/2,5/3)(41/63,86/63)(0.7,1.3)(1,1)
   };
    
    \addplot[
    name path=conv,
    thick,
    mark=square,]
    coordinates {
    (1/3,2)(1/2,5/3)(2/3,4/3)(1,1)
    };
    
   \legend{Known upper bound for $R^*(M)$, New lower bound for $R^*(M)$, Known lower bound for $R^*(M)$}
   
    \node[black,right] at (axis cs:1/2,1.68) {$(1/2,5/3)$};
    \node[black,left] at (axis cs:0.6,1.3) {$(2/3,4/3)$};
	\node[black,left] at (axis cs:0.6,1.4) {$(41/63,86/63)$};
	\node[black,left] at (axis cs:0.66,1.2) {$(0.7,1.3)$};
    \draw[->,>=stealth,very thick] (axis cs:0.6,1.4)--(axis cs:41/63,86/63);
    \draw[->,>=stealth,very thick] (axis cs:0.66,1.2)--(axis cs:0.7,1.3);
    \draw[->,>=stealth,very thick] (axis cs:0.6,1.3)--(axis cs:2/3,4/3);
    \end{axis}
	\end{tikzpicture}
	\caption{Rate-memory trade-off $R^*(M)$ and $R^*_L(M)$ for the $(3,3)$ cache problem.}
	\label{gen}
\end{figure}
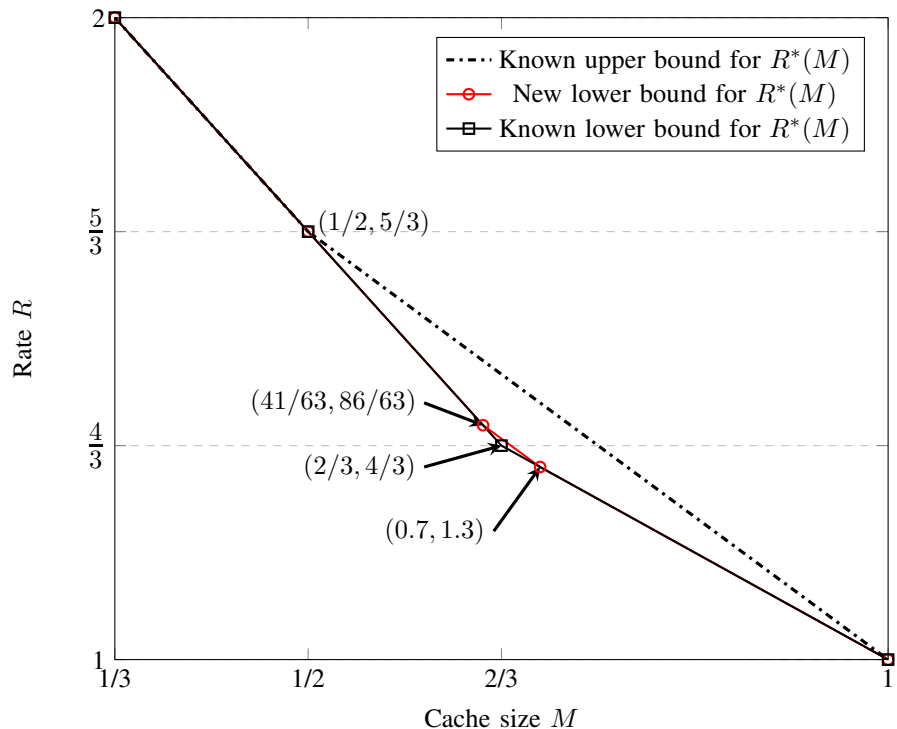

For the general case, we may use a variational version of the Ahlswede and K\"{o}rner Lemma \cite{kaced2013equivalence,makarychev2002new}.  Similarly to the linear case, we introduce an auxiliary random variable $G$, which satisfies the following constraint:
\begin{align}
    H(G|W_1X^{123})&=0,\\
    H(W_1|G)&=H(W_1|X^{213})\\
    H(X^{123}|G)&=H(X^{123}|X^{213})\\
    H(W_1X^{123}|G)&=H(W_1X^{123}|X^{213})
\end{align}
Then, by using a updated symmetry-reduced LP, we have:
\begin{equation}\label{eq:cz}
    41M+31R^*(M)\geq 69
\end{equation}
which slightly improve the lower bound of $R^*(M)$ (see Fig \ref{gen}). Moreover, this idea has been applied for the secret sharing in \cite{farras2018improving}.

\begin{remark}
This lower bound \eqref{eq:cz} shows that the point $(2/3,4/3)$ is not achievable and strengths the result that the point $(2/3,4/3)$ is not liner achievable in \cite{tian2018symmetry}.
\end{remark}

\bibliographystyle{IEEEtran}
\bibliography{cache}

\begin{thebibliography}{10}
\providecommand{\url}[1]{#1}
\csname url@samestyle\endcsname
\providecommand{\newblock}{\relax}
\providecommand{\bibinfo}[2]{#2}
\providecommand{\BIBentrySTDinterwordspacing}{\spaceskip=0pt\relax}
\providecommand{\BIBentryALTinterwordstretchfactor}{4}
\providecommand{\BIBentryALTinterwordspacing}{\spaceskip=\fontdimen2\font plus
\BIBentryALTinterwordstretchfactor\fontdimen3\font minus
  \fontdimen4\font\relax}
\providecommand{\BIBforeignlanguage}[2]{{%
\expandafter\ifx\csname l@#1\endcsname\relax
\typeout{** WARNING: IEEEtran.bst: No hyphenation pattern has been}%
\typeout{** loaded for the language `#1'. Using the pattern for}%
\typeout{** the default language instead.}%
\else
\language=\csname l@#1\endcsname
\fi
#2}}
\providecommand{\BIBdecl}{\relax}
\BIBdecl

\bibitem{maddah2014fundamental}
M.~A. Maddah-Ali and U.~Niesen, ``Fundamental limits of caching,'' \emph{IEEE
  Transactions on Information Theory}, vol.~60, no.~5, pp. 2856--2867, 2014.

\bibitem{tian2018symmetry}
C.~Tian, ``Symmetry, outer bounds, and code constructions: {A} computer-aided
  investigation on the fundamental limits of caching,'' \emph{Entropy},
  vol.~20, no.~8, p. 603, 2018.

\bibitem{cao2019coded}
D.~Cao, D.~Zhang, P.~Chen, N.~Liu, W.~Kang, and D.~G{\"u}nd{\"u}z, ``Coded
  caching with asymmetric cache sizes and link qualities: The two-user case,''
  \emph{IEEE Transactions on Communications}, vol.~67, no.~9, pp. 6112--6126,
  2019.

\bibitem{hammer2000inequalities}
D.~Hammer, A.~Romashchenko, A.~Shen, and N.~Vereshchagin, ``Inequalities for
  shannon entropy and kolmogorov complexity,'' \emph{Journal of Computer and
  System Sciences}, vol.~60, no.~2, pp. 442--464, 2000.

\bibitem{dougherty2009linear}
R.~Dougherty, C.~Freiling, and K.~Zeger, ``Linear rank inequalities on five or
  more variables,'' \emph{arXiv preprint arXiv:0910.0284}, 2009.

\bibitem{chen2016fundamental}
Z.~Chen, P.~Fan, and K.~B. Letaief, ``Fundamental limits of caching: {I}mproved
  bounds for users with small buffers,'' \emph{IET Communications}, vol.~10,
  no.~17, pp. 2315--2318, 2016.

\bibitem{gomez2016fundamental}
J.~G{\'o}mez-Vilardeb{\'o}, ``Fundamental limits of caching: Improved
  rate-memory tradeoff with coded prefetching,'' \emph{IEEE Transactions on
  Communications}, vol.~66, no.~10, pp. 4488--4497, Oct 2018.

\bibitem{yu2017characterizing}
Q.~Yu, M.~A. Maddah-Ali, and A.~S. Avestimehr, ``Characterizing the rate-memory
  tradeoff in cache networks within a factor of 2,'' \emph{arXiv preprint
  arXiv:1702.04563}, 2017.

\bibitem{tian2016symmetry}
C.~Tian, ``Symmetry, demand types and outer bounds in caching systems,'' in
  \emph{IEEE International Symposium on Information Theory (ISIT),}, 2016, pp.
  825--829.

\bibitem{kaced2013equivalence}
T.~Kaced, ``Equivalence of two proof techniques for non-shannon-type
  inequalities,'' in \emph{2013 IEEE International Symposium on Information
  Theory}.\hskip 1em plus 0.5em minus 0.4em\relax IEEE, 2013, pp. 236--240.

\bibitem{makarychev2002new}
K.~Makarychev, Y.~Makarychev, A.~Romashchenko, and N.~Vereshchagin, ``A new
  class of non-shannon-type inequalities for entropies,'' \emph{Communications
  in Information and Systems}, vol.~2, no.~2, pp. 147--166, 2002.

\bibitem{farras2018improving}
O.~Farr{\`a}s, T.~Kaced, S.~Mart{\'\i}n, and C.~Padr{\'o}, ``Improving the
  linear programming technique in the search for lower bounds in secret
  sharing,'' in \emph{Annual International Conference on the Theory and
  Applications of Cryptographic Techniques}.\hskip 1em plus 0.5em minus
  0.4em\relax Springer, 2018, pp. 597--621.

\end{thebibliography}
\end{document}